\documentclass[onecolumn,draftcls,12pt,twoside]{IEEEtranTCOM}

\normalsize

\usepackage{cite}
\usepackage{amsmath,amssymb,amsfonts}
\usepackage{algorithm}
\usepackage{algpseudocode}

\usepackage{graphicx}
  \usepackage{epstopdf}

\usepackage{textcomp}
\usepackage{bm}
\usepackage{xcolor}
\usepackage{lineno}

\ifCLASSOPTIONcompsoc
\usepackage[caption=false, font=normalsize, labelfont=sf, textfont=sf]{subfig}
\else
\usepackage[caption=false, font=footnotesize]{subfig}

\def\BibTeX{{\rm B\kern-.05em{\sc i\kern-.025em b}\kern-.08em
    T\kern-.1667em\lower.7ex\hbox{E}\kern-.125emX}}
\ifCLASSINFOpdf
\else
\fi

\begin{document}
%
\title{Asymmetric Quantum Concatenated and Tensor
Product Codes with   Large $Z$-Distances}
%
%
%

\author{Jihao~Fan,~\IEEEmembership{Member,~IEEE,}
        Jun~Li,~\IEEEmembership{Senior Member,~IEEE,}
        Jianxin~Wang, Zhihui~Wei and~Min-Hsiu Hsieh,~\IEEEmembership{Senior Member,~IEEE}
\thanks{J. Fan, J. Li, J. Wang, and Z. Wei are with School of Electronic and Optical Engineering, Nanjing University of Science and Technology,  Nanjing  210094,  China (e-mails: jihao.fan@outlook.com,     \{jun.li, wangjxin, gswei\}@njust.edu.cn). J. Fan is also with Key Laboratory of Computer Network and Information Integration (Southeast University), Ministry of Education, Nanjing 211189, China.}
\thanks{M.H. Hsieh  was with Center for Quantum Software and Information, Faculty of Engineering and Information Technology,
University of Technology Sydney, 15 Broadway, Ultimo NSW 2007, Australia, and is now     with Hon Hai Quantum Computing Research Center,   Taipei 114, Taiwan (email: minhsiuh@gmail.com)}
 }

%
%

\markboth{IEEE Transactions on Communications}%
{Submitted paper}
%



\maketitle
\begin{abstract}
In this paper, we present a  new construction of asymmetric quantum codes (AQCs)  by  combining classical concatenated  codes (CCs) with tensor product codes (TPCs), called asymmetric quantum concatenated and tensor product codes (AQCTPCs) which have  the following three advantages. First, only the outer codes in AQCTPCs need to satisfy the orthogonal constraint in quantum codes, and any classical linear code can be used for the inner, which makes AQCTPCs very easy to construct.
 Second, most AQCTPCs are highly degenerate, which means they can correct many more   errors than their classical TPC counterparts. Consequently, we construct  several families of AQCs with better parameters
than   known results in the literature.
Third, AQCTPCs can be efficiently decoded although they are   degenerate, provided that the inner and outer   codes
are efficiently decodable.    In particular, we significantly reduce the   inner  decoding complexity of TPCs from $\Omega(n_2a^{n_1})(a>1)$ to $O(n_2)$ by considering error degeneracy, where $n_1$ and $n_2$ are the block length of the inner code and the outer code, respectively.  Furthermore,   we generalize our
concatenation scheme by using the generalized CCs and TPCs correspondingly.
\end{abstract}


%
\IEEEpeerreviewmaketitle

\section{Introduction}
%
%
%
%
\newtheorem{definition}{Definition}[section]
\newtheorem{theorem}{Theorem}[section]
\newtheorem{lemma}{Lemma}[section]
\newtheorem{corollary}{Corollary}[section]
\newtheorem{example}{Example}[section]
\newtheorem{proposition}{Proposition}[section]
\label{sec:introduction}
\IEEEPARstart{Q}{uantum}    noise due to decoherence widely exists in quantum communication channels, quantum    gates and  quantum measurement. It is one of the biggest challenges  in realizing large-scale quantum communication systems and fully fault-tolerant quantum computation.  For a quantum state,  the two main mechanisms  of decoherence   are population relaxation and dephasing. The level of   noise is usually characterized by  the relaxation  time  $\mathcal{T}_1$ and the dephasing time $\mathcal{T}_2$. Further,  dephaisng usually generates a single  phase flip error, while population relaxation  generates  a mixed    bit-phase flip error.  It is shown in almost all quantum  systems that, the dephasing rate $1/\mathcal{T}_2$ is much faster than the relaxation rate $1/\mathcal{T}_1$, i.e., $\mathcal{T}_1\gg \mathcal{T}_2$ \cite{ioffe2007asymmetric,tuckett2018ultrahigh}. For example,  in the trapped
ions \cite{ioffe2007asymmetric,haffner2008quantum}, the ratio $\mathcal{T}_1/\mathcal{T}_2$ can
 be   larger than  $10^2$  and, in quantum dots systems \cite{churchill2009relaxation}, it can be larger than $10^4$.  Such large asymmetry between  population relaxation and dephasing indicates that phase flip errors ($Z$-errors) happen much more frequently than bit flip errors ($X$-errors).

Steane first saw that prior knowledge of this asymmetry in errors could be leveraged for performance gains and, hence, proposed asymmetric quantum codes (AQCs) in \cite{steane1996simple}. In the years since, many AQCs have been
  designed to have a biased error correction towards $Z$-errors  \cite{ioffe2007asymmetric,aliferis2008fault,sarvepalli2009asymmetric,brooks2013fault}.
For example,   AQCs constructed from   classical
Bose-Chaudhuri-Hocquenghem (BCH) codes \cite{macwilliams1981theory} and
low-density parity-check (LDPC) codes \cite{lin2004error,azmi2013ldpc,li2013design,yu2014generalized,li2011network}  were proposed in   \cite{ioffe2007asymmetric,sarvepalli2009asymmetric}. The BCH codes are used to correct $X$-errors and the more powerful LDPC codes are used to correct $Z $-errors. Another approach, devised by Galindo \emph{et al}. \cite{galindo2020asymmetric}, is to introduce some preshared entanglement \cite{brun2006correcting,hsieh2011high,wilde2013entanglement,hsieh2010entanglement,hsieh2009entanglement,fan2016constructions} to help construct AQCs. More recently, asymmetric errors have been explored as a way to help improve  the fault-tolerant  thresholds     \cite{aliferis2008fault,brooks2013fault}, particularly, in  topological quantum codes \cite{tuckett2018ultrahigh,PhysRevLett124130501,tuckett2019tailoring}.  In \cite{brooks2013fault}, a family of asymmetric Bacon-Shor (ABS) codes with  parameters $[[mn,1,m/n]]$,  where $m$ and $n$ are positive integers, is used for fault-tolerant quantum computation against highly biased noise. For example,    ABS codes with parameters $[[175,1,25/7]]$ and $[[315,1,35/9]]$ can achieve a very low  logical error rate  around $10^{-12}$ with much fewer physical two-qubit gates than symmetric quantum codes. In \cite{PhysRevLett124130501}, surface codes definied on a $d\times d$ square lattice of qubits  with $d=12,14,16,18,$ and $20$ have thresholds  exceeding $5\%$ when the asymmetry between $Z$-errors and $X$-errors is around $100$. Even more, it is shown recently in \cite{Bonilla2020} that thresholds for surface codes can exceed the zero-rate  Shannon bound of Pauli channels when the asymmetry is properly large! These results reveal  that
 the large asymmetry    in quantum channels has a significant effect to  quantum error correction and needs to be further exploited.

However, although there are many different constructions of AQCs in the literature, only a few
are made on binary AQCs with a relatively large $Z$-distance $d_Z$. This is because the dual-containing constraint in CSS codes often makes constructing an AQC with a  large minimum distance  $d_Z$ difficult. Aly  \cite{aly2008asymmetric} and Sarvepalli \emph{et al}. \cite{sarvepalli2009asymmetric} derived families of binary asymmetric quantum Bose-Chaudhuri-Hocquenghemmds (QBCH) codes  with  minimum distances $d_X$ and $d_Z$, both upper bounded by the square root of the block length.   Li \emph{et al.}   \cite{li2014two} were able to construct a few binary QBCH codes of length $n=2^m-1$ with a large minimum distance $d_Z$. Ezerman \emph{et al}. \cite{frederic2013csslike} constructed some binary CSS-like AQCs of length $\leq 40$ with best-known parameters by exhaustively searching the database of MAGMA  \cite{cannon2008handbook}.  Additionally, several families of nonbinary AQCs with a large   $d_Z$ have been developed, but all have a large field size   \cite{la2012asymmetric,galindo2017improved,la2014construction}.

The key to construct an AQC is to find two  classical  linear codes that satisfy a certain  dual-containing relationship. In classical codes, the two most useful combining  methods for constructing
linear codes from short constituent codes are: concatenated codes (CCs) \cite{forney1965concatenated} and tensor product codes (TPCs) \cite{wolf1965codes,wolf2006introduction}.    In general,   CCs have a large minimum distance because the distances in the constituent codes are multiplied, while TPCs have a poor minimum distance but a better dimension as a trade-off.
In \cite{maucher2000equivalence}, Maucher \emph{et al}.   show that generalized concatenated codes (GCCs) are equivalent to generalized tensor product codes (GTPCs).

It is not difficult to apply the concatenation method to the quantum realm, i.e., to construct concatenated quantum codes (CQCs) \cite{knill1996concatenated,gottesman1997stabilizer}  and quantum tensor product codes (QTPCs) \cite{grassl2005quantum,jihao2017on}, including asymmetric QTPCs   \cite{la2012asymmetric1} and entanglement-assisted QTPCs \cite{nadkarni2017entanglement}.
  CQCs and QTPCs also exhibit some similar characteristics to their classical counterparts. For example,   CQCs have a large minimum distance but a relatively small dimension, which is seeing them play an important role in fault-tolerant quantum computation.
   And, like TPCs, QTPCs have a large dimension   but a small minimum distance. However, it is worth noting that CQCs are not constructed from classical CCs directly, but rather by serially concatenating two constituent quantum codes. This means both the inner and outer constituent codes need to satisfy the dual-containing relationship, which limits their construction. The same does not apply to QTPCs, giving them a distinct advantage. But QTPCs usually  have a poor minimum distance. Moreover, some CQCs are known to be degenerate codes \cite{gottesman1997stabilizer}, which is a unique phenomenon in quantum coding theory. Degenerate codes have an advantage in that they can correct more errors than non-degenerate codes, but, in general, they are difficult  to decode (see \cite{hsieh2011np}) with the classical decoding algorithms often failing outright.

  Hence, in this paper, we propose a novel concatenation scheme called asymmetric quantum concatenated and tensor product codes (AQCTPCs) that combines both CCs and TPCs, where CCs are used to correct $Z$-errors, and TPCs are used to correct $X$-errors. Compared to the current methods, this new concatenation scheme has several advantages.
\begin{itemize}
\item [1)] In AQCTPCs, only the outer constituent codes over the extension field need to satisfy the dual-containing constraint. The inner constituent codes can be any classical linear codes. Then we have much freedom in the choice of the constituent codes.
\item[2)]  It is shown that AQCTPCs can be decoded efficiently provided that the classical constituent codes can be decoded efficiently.  In addition, AQCTPCs are highly degenerate for correcting $X$-errors and  they can correct many more $X$-errors beyond the error correction ability of the corresponding  TPCs.    Further, we show that the total inner decoding complexity  of TPCs is reduced significantly from $\Omega(n_2a^{n_1})(a>1)$ to $O(n_2)$ due to error degeneracy.   To this end, we have developed  a  syndrome-based decoding algorithm  specifically for AQCTPCs.
     \item[3)]
     The AQCTPCs demonstrated in this paper are    better than  QBCH codes or asymmetric quantum algebraic geometry (QAG) codes as the block length goes to infinity.  We   construct a family of AQCTPCs with a very large $Z$-distance  $d_Z$, of approximately half the block  length. Meanwhile, the dimension  and the $X$-distance  $d_X$ continue increasing as the block length goes to infinity. If $d_X=2$, then the $Z$-distance    $d_Z$ is larger than half the block length.
\end{itemize}
We compare  the parameters of AQCTPCs to previous results, and provide a generalized AQCTPC concatenation scheme that uses GCCs and GTPCs. We list     AQCTPCs with better parameters than the binary extension of asymmetric quantum Reed-Solomon (QRS) codes. We derive families of AQCTPCs with  the largest $Z$-distance  $d_Z$   compared to existed AQCs with comparable block length and $X$-distance  $d_X$.

The rest of this paper is organized as follows. In Section \ref{Preliminaries}, we provide the basic notations and definitions needed for the construction of AQCTPCs. In Section \ref{asymmetricquantumCCTPC}, we present the AQCTPC concatenation scheme and the decoding algorithms.  Section \ref{Families of AQCTPCs} provides detailed performance comparisons of AQCTPCs against previous constructions, and the discussions and conclusions follow in   Section \ref{disandconc}.

\section{Preliminaries}
\label{Preliminaries}
In this section we first review  some basic definitions and known results about   stabilizer codes and AQCs, followed by the introduction of classical CCs and TPCs and their generalizations.
\subsection{Stabilizer Codes and Asymmetric Quantum Codes}
 Denote by $q $   a power of a prime $p$ and denote by $\mathbb{F}_p$   the prime field. Let $\mathbb{F}_q$ be the finite field  with $q$ elements and let the   field $ \mathbb{F}_{q^m} $    be a field extension  of $\mathbb{F}_q$, where $m\geq1$ is an integer.   Let $\mathbb{C}$ be the field of complex numbers. For a positive integer $n$,
let $V_n=(\mathbb{C}^q)^{\otimes n}=\mathbb{C}^{q^n}$ be the $n$th tensor
product of $\mathbb{C}^q$. Denote by $u$ and $v$ two vectors of  $\mathbb{F}_q^n$. Define the error operators on $V_n$ by $X(u)|\varphi\rangle=|u+\varphi\rangle$ and $Z(v)|\varphi\rangle=\zeta^{Tr(v\varphi)}| \varphi\rangle$, where ``$Tr$'' stands for the trace operation from $\mathbb{F}_q$ to $\mathbb{F}_p$, and $\zeta=\exp(2\pi i/p)$ is a primitive $p$th root of unity. Denote by
\begin{equation}
G_n=\{\zeta^a X(u)Z(v):  u,v\in\mathbb{F}_q^n, a\in \mathbb{F}_q\}
\end{equation}
the   group generated by $E_n=\{X(u)Z(v):u,v\in\mathbb{F}_q^n\}$. For any $\varepsilon=\zeta^aX(u)Z(v)\in G_n$, where $u=(u_1,\ldots,u_n)\in\mathbb{F}_q^n$ and $v=(v_1,\ldots,v_n)\in\mathbb{F}_q^n$, the   weight of $\varepsilon$ is defined by \begin{equation}
\textrm{wt}_Q(\varepsilon)=|\{1\leq i\leq n: (u_i,v_i)\neq(0,0)\}|.
 \end{equation}
 The weight of $X$-errors and the weight of $Z$-errors in $\varepsilon$ are defined by $\textrm{wt}_H(u)$ and $\textrm{wt}_H(v)$, respectively, where ``$\textrm{wt}_H $'' stands for the Hamming weight. The definition of quantum stabilizer codes is given below.
\begin{definition}
\label{definition of QEC and AQC}
A $q$-ary quantum stabilizer code $\mathcal{Q}$ is a $q^k$-dimensional ($k>0$) subspace of $V_n$ such that
 \begin{equation}
\mathcal{Q}=\bigcap_{\varepsilon\in S}\{|\varphi\rangle\in V_n: \varepsilon|\varphi\rangle=|\varphi\rangle\},
\end{equation}
where $S$ is a subgroup of $G_n$ and is called the stabilizer group.  $\mathcal{Q}$ has minimum distance $d$ if   it
can detect all  errors  $ \varepsilon\in G_n$ of  weight $\textrm{wt}_Q(\varepsilon)$ up to $d-1$. Then $\mathcal{Q}$ is denoted by $\mathcal{Q}=[[n,k,d]]_q$.  Further, $\mathcal{Q}$ is called non-degenerate if each    stabilizer in $S$  has quantum weight   at least the minimum distance $d$, otherwise it is degenerate.\end{definition}

The Calderbank-Shor-Steane (CSS) code  in \cite{calderbank1996good,steane1996simple} is a special family of quantum stabilizer codes and can be constructed  from two classical linear codes which satisfy some dual-containing relationship.  Let $d_X$ and $d_Z$ be two positive integers.  We define an AQC as a CSS code  in $V_n$ with parameters $Q=[[n,k,d_Z/d_X]]_q$    if it can
detect all    $ \varepsilon\in G_n$ of   weight $\textrm{wt}_X(\varepsilon)$ up to  $d_X-1$   and  weight $\textrm{wt}_Z(\varepsilon)$ up to  $d_Z-1$, simultaneously.
The   construction in   \cite{sarvepalli2009asymmetric,ioffe2007asymmetric,wang2010asymmetric} can be used to construct AQCs in which a pair of classical linear codes are used, one for correcting   $X$-errors and
the other  for correcting   $Z $-errors.

\begin{lemma}[{\cite[Theorem 2.4]{wang2010asymmetric}}]
\label{AQC Constructions}
Let $C_1$ and $C_2$ be two classical
linear codes with parameters
$[n,k_1,d_1]_q$ and $[n,k_2,d_2]_q$, respectively, and   $C_2^{\bot}\subseteq C_1$. Then there exists an AQC with parameters $ {Q}=[[n,k_1+k_2-n,d_Z/ d_X]]_q,$ where
\begin{equation}
d_Z=\max\{\text{wt}_H(C_1\backslash C_2^{\bot}),\text{wt}_H(C_2\backslash C_1^{\bot})\},
\end{equation}
\begin{equation}
d_X=\min\{\text{wt}_H(C_1\backslash C_2^{\bot}),\text{wt}_H(C_2\backslash C_1^{\bot})\}.
\end{equation}
If $d_1= \text{wt}_H(C_1\backslash C_2^{\bot})$ and $d_2=\text{wt}_H(C_2\backslash C_1^{\bot})$, then $Q$ is non-degenerate, otherwise it is degenerate.
\end{lemma}

\subsection{Classical Tensor Product Codes}

Let $C_1=[n_1, k_1, d_1]_q$ be   a classical linear code whose parity check matrix is given by $H_{c_1}$,   and let $r_1= n_1- k_1$
be the number of  parity  checks.    Let  $C_2=[n_2, k_2, d_2]_{q^{r_1}}$  be a linear code over the  extension field
$ \mathbb{F}_{q^{r_1}} $  whose parity check matrix is given by  $H_{c_2}$. Let $r_2=n_2-k_2$.  Denote by
\begin{equation}
 C_T \equiv C_2\otimes_TC_1
\end{equation}
the tensor product code  of $C_1$ and $C_2$. The block length and dimension of $C_T$ are given by $ [n_1n_2,n_1n_2-r_1r_2]$. In addition, $C_1$ and $C_2$ are known as  the inner and outer constituent codes of $C_T$, respectively. If  we
regard  $H_{c_1}$  as a $1\times n_1$ matrix with elements over $ \mathbb{F}_{q^{r_1}}$,  then the parity check matrix $H_T$ of   $C_T$ is  the Kronecker  product of $H_{c_1}$ and $H_{c_2}$, i.e., 
\begin{equation}
H_T = H_{c_2}\otimes H_{c_1}.
\end{equation}
 Then we can derive a parity check matrix of $C_T$ with elements over $ \mathbb{F}_{q }$ by expanding all the elements of $H_T$ from $ \mathbb{F}_{q^{r_1}}$ to $ \mathbb{F}_{q}$. The error  detection/correction ability of $C_T$ is restricted  by  the constituent codes and is given by:
\begin{lemma}[{\cite[Theorem 1]{wolf1965codes}}]
\label{wolftheorem}
Partition the codeword of   $C_T = C_2\otimes_TC_1$ into $n_2$
sub-blocks, where each sub-block contains  $n_1$ elements, and assume that the constituent code  $C_i$  can detect or correct an error pattern class $\xi_i$ ($i=1$ or 2),   then
the TPC $C_T $ can detect or correct all error-patterns where the
sub-blocks containing errors form a pattern belonging to
class  $\xi_2$ and the errors within each erroneous sub-block fall
within the class $\xi_1$.
\end{lemma}

 Here we give an  illustrative example for the construction of TPCs.
 \begin{example}
 Let $C_1=[3,1,3]_2$ be a binary repetition code with a parity check matrix given by
 \begin{eqnarray}
 H_{c_1}&=&\left(\begin{array}{ccc}
 1 &0& 1 \\
 0 &1 &1
 \end{array}\right)=\left(\begin{array}{ccc}
  1&\omega&\omega^2
 \end{array}\right),
 \end{eqnarray}
 where $\omega$ is a primitive element of $GF(2^2)$ such that $\omega^2+\omega+1=0$. Let $C_2$ be a $2^2$-ary linear code over $GF(2^2)$, such as we let $C_2=[5,3,3]_{2^2}$ be a   maximum-distance-separable (MDS) code with a parity check matrix
 \begin{equation}H_{c_2}=\left(\begin{array}{ccccc}
  1&0&1&\omega&\omega \\
  0&1&\omega&\omega &1
 \end{array}\right).\end{equation}   Then   we can derive a TPC   $C_T $ of length $15$ whose   parity check matrix $H_T=H_{c_2}\otimes H_{c_1}$  is  given in (\ref{HTbinary}).
 \newcounter{mytempeqncnt}
\begin{figure*}[t]
 \begin{eqnarray}\nonumber
H_T&=&\left(\begin{array}{ccccccccccccccc}
  1&\omega&\omega^2& 0&0&0& 1  &\omega&\omega^2& \omega&\omega^2&1& \omega&\omega^2&1 \\
  0&0&0& 1&\omega&\omega^2& \omega &\omega^2&1& \omega&\omega^2&1& 1&\omega&\omega^2
 \end{array} \right)  \\ \label{HTbinary}
&=&\left(\begin{array}{ccccccccccccccc}
  1&0&1& 0&0&0& 1 & 0&1&0& 1&1&0& 1  &1  \\
  0&1&1& 0&0&0& 0  &1&1&1& 1&0&1& 1 &0 \\
   0&0&0& 1&0&1& 0 & 1&1&0& 1&1&1& 0  &1 \\
  0&0&0& 0&1&1& 1 &1&0&1& 1&0&0& 1 &1
 \end{array} \right) .
 \end{eqnarray}
\hrulefill
\end{figure*}
 It is easy to verify,  e.g., by using the MAGMA computational software  \cite{cannon2008handbook}, that the dimension and minimum distance of $C_T$ with a parity check matrix $H_T $ in (\ref{HTbinary}) are exact $11$ and $3$, respectively.
 \end{example}

Ref. \cite{maucher2000equivalence} shows that the  parity check  matrix of TPCs can also be    represented   in a   companion matrix form.  Let $g(x)=g_0+g_1x+\cdots+g_{r_1-1}x^{r_1-1}+x^{r_1}$ be a primitive polynomial over $\mathbb{F}_{q^{r_1}}$ and denote by $\alpha$ a primitive element of $\mathbb{F}_{q^{r_1}}$. The companion matrix  of $g(x)$ is defined to be the $r_1\times r_1$ matrix
  \begin{equation}
\label{companion_matrix}
M =
\left(
\begin{array}{ccccc}
 0&1&0&\cdots&0\\
 0&0&1&\cdots&0\\
\vdots&\vdots&\vdots&\ddots&\vdots\\
 0&0&0&\cdots&1\\
-g_0&-g_1&-g_2&\cdots&-g_{r_1-1}
\end{array}
\right) .
\end{equation}
Then for any element $\beta=\alpha^i$ of $\mathbb{F}_{q^{r_1}}$, the companion matrix of $\beta$, denoted by $[\beta]=M^i$, is an $r_1\times r_1$ matrix with elements over $ \mathbb{F}_q$.
Let the parity check matrix of the constituent  code $C_2$ be  $H_{c_2}=(a_{ij})_{r_2\times n_2}$   with  elements over $\mathbb{F}_{q^{r_1}}$, i.e., $a_{ij}\in  \mathbb{F}_{q^{r_1}}$ for $1\leq i \leq r_2$ and $1\leq j\leq n_2$. Following the notations used in \cite{maucher2000equivalence}, we denote by $[H_{c_2}]=([a_{ij}])_{r_1r_2\times r_1n_2}$,  where $[a_{ij}]$ is a companion matrix form.
The parity check matrix of $C_T$ can   be written as
\begin{eqnarray}
\label{companion_matrix_TPC1}
\nonumber
 \hspace{-7mm}&&H_T \equiv [H_{c_2}^t]\otimes  H_{c_1} \\
\hspace{-7mm}&&= \left(
\begin{array}{cccc}
[a_{11}^t]H_{c_1}& [a_{12}^t]H_{c_1}&\cdots& [a_{1n_2}^t]H_{c_1} \\

[a_{21}^t]H_{c_1}& [a_{22}^t]H_{c_1}&\cdots& [a_{2n_2}^t]H_{c_1} \\
\vdots&\vdots&\vdots&\vdots\\

[a_{r_{21}}^t]H_{c_1}& [a_{r_{22}}^t]H_{c_1}&\cdots& [a_{r_2n_2}^t]H_{c_1}
\end{array}
\right)
\end{eqnarray}
in which the matrix $[H_{c_2}^t]$ is obtained by transposing  the constituent  companion matrices  of $[H_{c_2} ]$, and $[a_{ij}^t]$ is the transpose of $[a_{ij}]$.
According to \cite{maucher2000equivalence,fan2017comments}, if we do not transpose the constituent companion  matrices in (\ref{companion_matrix_TPC1}), we  can obtain another representation of the parity check matrix $H_T $ as follows
\begin{eqnarray}
\label{companion_matrix_TPC2}
\nonumber
 \hspace{-7mm}&&H_T \equiv  [H_{c_2}]\otimes  H_{c_1} \\
\hspace{-7mm}&&= \left(
\begin{array}{cccc}
[a_{11}]H_{c_1}& [a_{12}]H_{c_1}&\cdots& [a_{1n_2}]H_{c_1} \\

[a_{21}]H_{c_1}& [a_{22}]H_{c_1}&\cdots& [a_{2n_2}]H_{c_1} \\
\vdots&\vdots&\vdots&\vdots\\

[a_{r_{21}}]H_{c_1}& [a_{r_{22}}]H_{c_1}&\cdots& [a_{r_2n_2}]H_{c_1}
\end{array}
\right).
\end{eqnarray}
The two   representations  in (\ref{companion_matrix_TPC1}) and (\ref{companion_matrix_TPC2}) do not make any difference for the parameters and the error correction performance of TPCs. We will use   them alternately in the following constructions. It should be noticed that the Kronecker product defined in equations (\ref{companion_matrix_TPC1}) and (\ref{companion_matrix_TPC2}) is a little different from the standard one. In the following,   the Kronecker product of matrices    follows  the definition in   (\ref{companion_matrix_TPC1}) and (\ref{companion_matrix_TPC2}).

The generalized tensor product codes  are proposed  in \cite{maucher2000equivalence,imai1981generalized} by combining a series of outer codes and  inner codes. Let $A_\hbar=[n_A,k_{\hbar},d_{\hbar}]_{q}$ and    $B_\hbar=[N_B,K_{\hbar},D_{\hbar}]_{q^{r_\hbar}}$ be $L$ pairs of inner and outer codes, respectively,  where  $1\leq\hbar\leq L$ and $r_\hbar=n_A-k_\hbar$. Let the parity check matrices of $A_\hbar $ and    $B_\hbar $, respectively,  be $H_{\hbar}^A$ and $H_{\hbar}^B$, $1\leq\hbar\leq L$. Assume that all the rows in $H_\hbar^A$, $1\leq\hbar\leq L$, are independent with each other.  Then the parity check matrix of the GTPCs
\begin{equation}
\mathcal{C}_{\mathcal{T}}=\bigcap\limits_{\hbar=1}^{L} B_\hbar\otimes_TA_\hbar
\end{equation}
 is defined by
\begin{equation}
H_{\mathcal{C}_{\mathcal{T}}}\equiv\left(
\begin{array}{c}
[H^{B^t}_1]\otimes H^A_1\\

[H^{B^t}_2]\otimes H^A_2\\
\vdots\\

[H^{B^t}_L]\otimes H^A_L
\end{array}
\right),
\end{equation}
where $[H^{B^t}_\hbar]$ is   obtained  by transposing the component companion matrices of $[H^{B}_\hbar]$ for each $1\leq\hbar \leq L$. The block length and the dimension of GTPCs are given by $\mathcal{C}_{\mathcal{T}}=[N_Bn_A,N_Bn_A-\sum_{\hbar=1}^LR_\hbar r_\hbar ]_q$, where $R_\hbar=N_B-K_\hbar$ for $ 1\leq \hbar\leq L$.

\subsection{Classical Concatenated Codes}

Concatenated codes can be seen as the dual counterpart  of TPCs, which are obtained by concatenating an inner code $C_1=[n,k,d]_q$ with an outer code $C_2=[N,K,D]_{q^k}$. Denote   the concatenation of $C_1$ and $C_2$ by
\begin{equation}
C_C\equiv C_2\otimes_C C_1,
\end{equation}
and  $C_C=[ Nn, Kk, d_{C_C}\geq Dd]_q$ (see \cite{macwilliams1981theory,forney1965concatenated}). The generator matrix of $C_C$   can  also  be given in a companion matrix form  (see \cite{maucher2000equivalence})
\begin{equation}
G_C=[G_2]\otimes G_1.
\end{equation}
where $G_1$ and $G_2$ are the  generator matrices of $C_1$ and $C_2$, respectively.

 In \cite{macwilliams1981theory,maucher2000equivalence}, the generalized concatenated codes    are  obtained by concatenating a serial of outer codes and inner codes. For simplicity, we only consider linear codes here.  Let $A_1=[n_A,k_{1},d_{1}]_q$ be a $q$-ary linear code with the generator matrix $G_1^A$, which is partitioned to $S $ submatrices $ \mathbf{G}_1^A,\ldots, \mathbf{G}_S^A$ such that $k_\ell^A=\textrm{rank}(\mathbf{G}_\ell^A)$ for $1\leq\ell\leq S$, and then $k_1=\sum_{\ell=1}^{S}k_\ell^A$. Denote by
\begin{equation}
\label{G1A}
G_1^A=\left(
\begin{array}{c}
 \mathbf{G}_1^A\\

 \mathbf{G}_2^A\\
\vdots\\

 \mathbf{G}_S^A
\end{array}
\right),
G_\ell^A=\left(
\begin{array}{c}
 \mathbf{G}_\ell^A\\

 \mathbf{G}_{\ell+1}^A\\
\vdots\\

 \mathbf{G}_S^A
\end{array}
\right),2\leq\ell\leq S,
\end{equation}
 and let $G_\ell^A$ be the generator matrices of the linear codes   $A_\ell=[n_A,k_{\ell},d_{\ell}]_q$, for $2\leq\ell\leq S$, respectively.   Denote by $B_\ell=[N_B,K_{\ell},D_{\ell}]_{q^{k_\ell^A}}$   the outer codes with the generator matrices, respectively, $G_\ell^B$,  for $1\leq\ell\leq S$.   Then the generator matrix of the GCCs
\begin{equation}
\mathcal{C}_{\mathcal{C}}=\bigcup\limits_{\ell=1}^{S} B_\ell\otimes_CA_\ell
\end{equation}
is defined by
\begin{equation}
G_{\mathcal{C}_\mathcal{C}}\equiv\left(
\begin{array}{c}
[G^{B}_1]\otimes \mathbf{G}^A_1\\

[G^{B}_2]\otimes \mathbf{G}^A_2\\
\hspace{2mm}\vdots\\

[G^{B}_S]\otimes \mathbf{G}^A_S
\end{array}
\right),
\end{equation}
and the parameters of GCCs are given by
\begin{equation}
\mathcal{C}_{\mathcal{C}}=[N_Bn_A,\sum\limits_{\ell=1}^SK_\ell k_\ell^A,d_{\mathcal{C}_{\mathcal{C}}}]]_q,
\end{equation}
where $d_{\mathcal{C}_{\mathcal{C}}}\geq \min\{D_1d_1 ,\ldots,D_Sd_S\}$.

Compared to   other types of classical linear codes in \cite{macwilliams1981theory,Grassl:codetables}, the parameters of CCs (GCCs) and TPCs (GTPCs) may not have any advantages. However the encoding and decoding algorithms of CCs (GCCs) and TPCs (GTPCs) usually have low complexity, and can be decoded efficiently in polynomial time. Therefore
CCs are widely used in many digital communication systems, e.g., the NASA standard for deep space communications and wireless communications \cite{costello2007channel,lin2004error}, and GCCs show   large potential applications, e.g.,  in data
transmission systems \cite{zhilin2015high} and Flash memory \cite{spinner2016soft,rajab2018soft}.   TPCs and GTPCs exhibit large advantages in magnetic storage systems \cite{alhussien2010iteratively, chaichanavong2006tensor, chaichanavong2006tensor1,fahrner2004low}, Flash memory \cite{gabrys2013graded,kaynak2014classification} and in constructing      locally repairable codes for distributed storage systems \cite{huang2015linear,huang2015binary,blaum2019extended}.   In \cite{trifonov2016fast}, it is shown that Polar codes can be treated as GCCs for a fast encoding.
\section{Main Results}
 \label{asymmetricquantumCCTPC}
In this section, we present the AQCTPC concatenation framework, where CCs are used to correct $Z$-errors and TPCs are used to correct $X$-errors. In our construction, the dimension of the inner   codes of   CCs needs to be equal to the number of parity checks of the inner   codes of TPCs.
Let  $C_1=[n_1,k_1,d_1]_q $  denote    an arbitrary $q$-ary linear code and    $C_2=[n_2,k_2,d_2]_{q^{k_1}}$  and $C_3=[n_2,k_3,d_3]_{q^{k_1}}$  denote two linear codes  over the extension field $ \mathbb{F}_{q^{k_1}}$.  Let $\mathcal{C}_C=C_3\otimes_C C_1$ be the CC of $C_1$ and $C_3$, and let $\mathcal{C}_T=C_2\otimes_T C_1^\bot$ be the TPC of $C_1^\bot$ and $C_2$. Then we have the following dual-containing relationship between CCs and TPCs.
\begin{lemma}
\label{asymmetric tensor and concatenated}
 If $C_3^\bot\subseteq C_2$, then there exists     $\mathcal{C}_T^\perp\subseteq\mathcal{C}_C$.
\end{lemma}
\begin{IEEEproof}
 Let   $H_{c_1}$  and $G_{c_1}$ be   the parity check matrix and generator matrix   of $C_1  $ over $\mathbb{F}_q$, respectively. Let   $H_{c_i}$  and $G_{c_i}$,  $i=2,3$, be   the parity check matrix and generator matrix of $C_i$ over $ \mathbb{F}_{q^{k_1}}$, respectively. It is easy to see that the parity check matrix  of the TPC $\mathcal{C}_T$ with transposed companion matrices    is given by
 \begin{equation}
  \label{pariy-check-X}
 H_{\mathcal{C}_T}= [H_{c_2}^t]\otimes  G_{c_1}.
 \end{equation}
 From \cite{maucher2000equivalence,fan2017comments}, we know that the  parity check matrix of  $\mathcal{C}_C$ is given by
  \begin{equation}
H_{\mathcal{C}_C}= \left(
            \begin{array}{l}
             [H_{c_3}]\otimes  (I_{k_1},\textbf{0})\\

            [I_{n_2}]  \otimes \  H_{c_1}
            \end{array}
       \right),
 \end{equation}
 where ``$\textbf{0}$'' is a zero sub-block of size $k_1\times (n_1-k_1)$.
  It is not difficult to verify that if $C_3^\bot\subseteq C_2$, then we have $[H_{c_3}][H_{c_2}^t]^T=0$ and  $H_{\mathcal{C}_C}H_{\mathcal{C}_T}^T=0$. Therefore we have $\mathcal{C}_T^\perp\subseteq\mathcal{C}_C$.
\end{IEEEproof}

By combining     $\mathcal{C}_C=C_3\otimes_C C_1$ and     $\mathcal{C}_T=C_2\otimes_T C_1^\bot$, we have the construction of AQCTPCs as follows.

\begin{theorem}
\label{AQCTPCTheorem}
There exists a family of  AQCTPCs with parameters $\mathcal{Q}=[[n_1n_2,k_1(k_2+k_3-n_2),d_Z\geq d_1d_3/d_X\geq d_2]]_q$.
\end{theorem}

\begin{figure}
  \centering
  \includegraphics[width=3.2in]{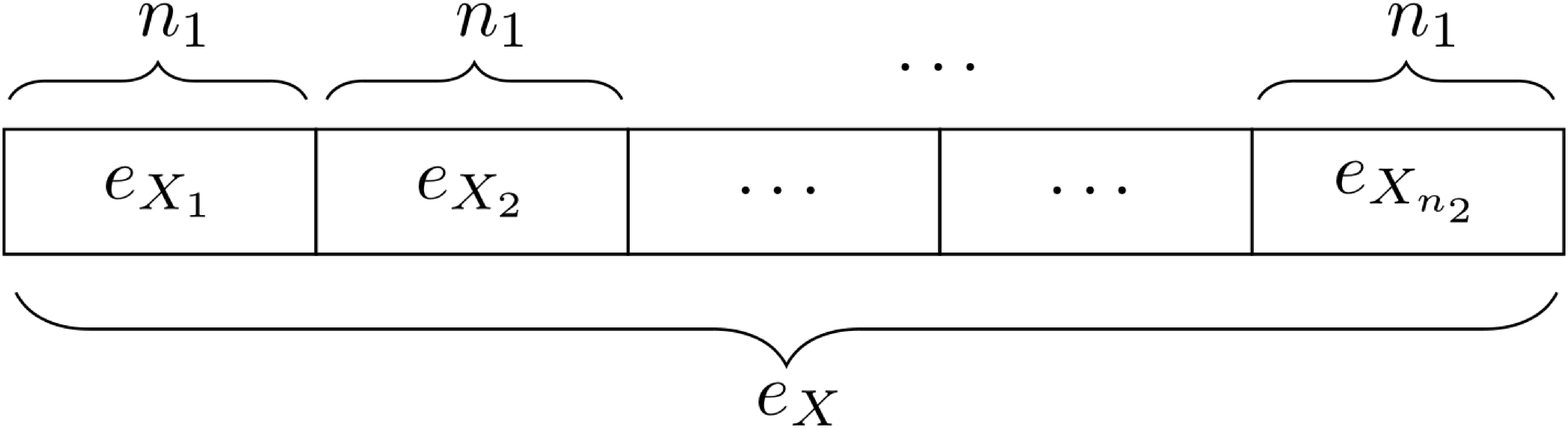}\\
  \caption{Dividing the Pauli $X$-error $e_X$ into $n_2$ sub-blocks where each sub-block  $e_{X_i} (1\leq i\leq n_2)$ is of length $n_1$.}\label{sublocks}
\end{figure}

 The AQCTPC concatenation scheme has several advantages over the current methods. First,    only the outer constituent codes $C_2$ and $C_3$ over the extension field need  to satisfy the  dual-containing constraint.  Then we have much freedom  in the choice of the outer  codes.  It is generally believed that  certain families of linear  codes over the extension field  can easily satisfy the dual-containing constraint. For example, the dual-containing relationship of $\ell$-ary     MDS codes has been   determined for all possible dimensions and block length less than $\ell+2$, see e.g., in \cite{grassl1999quantum,grassl2004optimal,ezerman2013pure}, where $\ell$ is   a power of a prime.   We can let $C_2$ and $C_3$  be  two MDS codes  that satisfy the
dual-containing  constraint and with reasonable block length.  Second,    in the following proof we show that AQCTPCs are highly degenerate in that they can correct more $X$-errors than a corresponding classical TPC.
\begin{IEEEproof}[Proof of Theorem \ref{AQCTPCTheorem}]
Let $\mathcal{C}_C=C_3\otimes_C C_1$ denote the  CC   of $C_1$ and $C_3$, and let    $\mathcal{C}_T=C_2\otimes_T C_1^\bot$ denote the TPC of $C_1^\bot$ and $C_2$.  Then we have $\mathcal{C}_C=[n_1n_2,k_1k_2]_q$  and $\mathcal{C}_T =[n_1n_2,n_1n_2-k_1(n_2-k_2)]_q$.
According to the CSS construction in Lemma \ref{AQC Constructions} and Lemma \ref{asymmetric tensor and concatenated}, if $C_3^\bot\subseteq C_2$, then we can derive an AQCTPC  with parameters
$
\mathcal{Q}=[[n_1n_2,k_1(k_2+k_3-n_2),d_Z/d_X]]_q.
$

We still need to compute  the minimum distance of $\mathcal{Q}$.
It is easy to see that the minimum distance of   the CC   is larger than or equal to $d_1d_3$, and then we have $d_Z\geq d_1d_3$.  Next we determine the $X$-distance  $d_X$.

Suppose that there is an   $X$-error $e_X$ of length $n_1n_2$  in the encoded codeword. We divide the error $e_X$   into $n_2$ sub-blocks   $e_{X_i}(1\leq i\leq n_2)$, with each sub-block  being of length $n_1$ (see Fig. \ref{sublocks}).
We then do the syndrome measurement for $X$-errors by using the parity check matrix $H_{\mathcal{C}_T}$ given in (\ref{pariy-check-X}).    The syndrome information $\Phi $    can be derived  by measuring the ancilla, which is given by
\begin{eqnarray}
\nonumber
 \Phi&\equiv & [H_{c_2}^t]\otimes  G_{c_1}\cdot e_X^T \\
 \hspace{-7mm}& =&\left(
\begin{array}{c}
[a_{11}^t]\Phi_{\textrm{I}_1}+\cdots+[a_{1n_2}^t]\Phi_{\textrm{I}_{n_2}} \\

[a_{21}^t]\Phi_{\textrm{I}_1}+\cdots+ [a_{2n_2}^t]\Phi_{\textrm{I}_{n_2}} \\
\hspace{-3mm}\vdots \\

[a_{r_21}^t]\Phi_{\textrm{I}_1}+\cdots+ [a_{r_2n_2}^t]\Phi_{\textrm{I}_{n_2}}
\end{array}
\right),
\end{eqnarray}
 where $H_{c_2}=(a_{ij}), 1\leq i\leq r_2=n_2-k_2, 1\leq j\leq n_2 $.   We let
 $
 \Phi_{\textrm{I}_l}\equiv G_{c_1}  e_{X_l}^T, 1\leq l\leq n_2,
$
 which can be regarded as the logical error sequences in the outer code $C_2$. Then we have
\begin{equation}
 \Phi=\left(
\begin{array}{cccc}
[a_{11}^t]  &[a_{12}^t]  &\cdots& [a_{1n_2}^t]  \\

[a_{21}^t]  &[a_{22}^t]  &\cdots& [a_{2n_2}^t]  \\
\vdots&\vdots&\vdots&\vdots\\

[a_{r_21}^t] &[a_{r_22}^t] &\cdots& [a_{r_2n_2}^t]
\end{array}
\right)\left(\begin{array}{c}
  \Phi_{\textrm{I}_1} \\

\Phi_{\textrm{I}_2} \\
\vdots\\

\Phi_{\textrm{I}_{n_2}}
\end{array}\right).
\end{equation}
 If the   outer decoding  can be conducted successfully, then the sequences $ \Phi_{\textrm{I}_l}(1\leq\ell \leq n_2)$ are used as the inner syndrome information for  $ C_1^\bot$.

The outer code $C_2$ must be decoded by mapping the syndrome information $\Phi$ to the symbols over the extension field  $\mathbb{F}_{q^{k_1}}$. Here we need a syndrome based decoding \cite{macwilliams1981theory} of the outer code $C_2$, which, if successful, will result in the exact inner syndrome sequence  $\Phi_{\textrm{I}_l}(1\leq l\leq n_2)$.  The inner decoding follows using the dual of the inner code $C_1$.  In general, for any $ \Phi_{\textrm{I}_l}\equiv G_{c_1}  e_{X_l}^T( 1\leq l\leq n_2 )$, we can always  obtain   a decoded error sequence  $\widetilde{e}_{X_l}$ such that $\Phi_{\textrm{I}_l}=G_{c_1}  \widetilde{e}_{X_l}^T$  by using some syndrome based decoder for $C_1^\bot$, such as a  syndrome table-look-up decoder. However, we do not need to do that   and  just let $\widetilde{e}_{X_l}=(\Phi_{\textrm{I}_l},\textbf{0}) $ by assuming that $ G_{c_1}$ is in a standard form, where ``$\textbf{0}$'' is a zero vector of length $ r_1$. Let $\widetilde{e}_X\equiv(\widetilde{e}_{X_1},\ldots, \widetilde{e}_{X_{n_2}}) $ be the decoded error sequence. There must be $G_{\mathcal{C}_C}(e_X^T+\widetilde{e}_X^T)= 0$,   where
\begin{equation}
G_{\mathcal{C}_C}=   [G_{c_3}^t]\otimes  G_{c_1}
 \end{equation}
 is the generator matrix of $\mathcal{C}_C=C_3\otimes_C C_1$. There are two cases: (1) $e_X =\widetilde{e}_X $ which means that the decoded error  $\widetilde{e}_X $ is exactly the true error. (2)    $e_X \neq\widetilde{e}_X $ but they belong to the same coset of  $ \mathcal{C}_C^\bot$, which means that they are degenerate.

 This phenomenon of degeneracy is quite different from the decoding of classical TPCs  \cite{alhussien2010iteratively,wolf1965codes,maucher2000equivalence}, where the decoding fails if the number of errors in one sub-block exceeds the error correction ability of the inner codes. As such, AQCTPCs can correct many more $X$-errors than their classical TPC counterparts.
 If $
 \textrm{wt}(e_X)\leq d_2-1
$, whenever the error is separated into different sub-blocks in Fig. \ref{sublocks}, the number of erroneous sub-blocks will be at most $d_2-1$. This means that either the error will always be detectable or that the error is undetectable but harmless since it is degenerate. Thus  the $X$-distance   $d_X$ is at least $d_2$. Therefore we have an AQCTPC with the parameters $\mathcal{Q}=[[n_1n_2,k_1(k_2+k_3-n_2),d_Z\geq d_1d_3/d_X\geq d_2]]_q$.
   \end{IEEEproof}

In the proof of Theorem \ref{AQCTPCTheorem}, we have given the decoding of AQCTPCs for correcting $X$-errors.
 We summarize and provide the whole decoding   process in Algorithm \ref{DecTPCs}.
\renewcommand{\algorithmicrequire}{\textbf{Input:}}
\renewcommand{\algorithmicensure}{\textbf{Output:}}
\begin{algorithm}[H]
\caption{The Decoding Algorithm of AQCTPCs for Correcting $X$-errors.}
\label{DecTPCs}
\begin{algorithmic}[1]
\Require
  $\Phi,
  H_{c_2}$;
\Ensure The decoded $X$-error sequence $\widetilde{e}_X$.
\State            Initialization: $\widehat{\Phi} = \emptyset$, $\widetilde{e}_{X}=\emptyset$;
\State $//$ \verb"Divide" $\Phi$ \verb"into" $r_2$ \verb"sub-blocks, each" \verb"sub-block is of length "$k_1$.
\State  $\Phi =(\Phi_1,\ldots,\Phi_{r_2}),  |\Phi_i|= k_1$;
\State $//$ \verb"Map" $\Phi$ \verb"to" $\widehat{\Phi}$ \verb"with elements over" \verb"the extension field "$\mathbb{F}_{q^{k_1}}$.
\For{  $i\in [1,r_2]$}
\State Map $\Phi_i$ into a symbol $\widehat{\Phi}_i$ over the field $\mathbb{F}_{q^{k_1}}$;
\State  $\widehat{\Phi}=(\widehat{\Phi},\widehat{\Phi}_i) $;
\EndFor
\State $//$ \verb|Do the outer decoding according| \verb|to the syndrome information| $H_{c_2}\widehat{\Phi}_{\textrm{I}}^T =\widehat{\Phi}$.
 \State Denote by $\widehat{\Phi}_\textrm{I}=(\widehat{\Phi}_{\textrm{I}_1},\ldots,\widehat{\Phi}_{\textrm{I}_{n_2}})$;
 \For{  $i\in [1,n_2]$}
\State Map  $\widehat{\Phi}_{\textrm{I}_i}$ into a sequence over field   $\mathbb{F}_{q}$,  $ {\Phi}_{\textrm{I}_i}$;
\State $\widetilde{e}_{X_i}=(\Phi_{\textrm{I}_i},\textbf{0}) $;
\State  $\widetilde{e}_{X}=(\widetilde{e}_{X},\widetilde{e}_{X_i})$;
\EndFor  \\
\Return $\widetilde{e}_{X}$;
\end{algorithmic}
\end{algorithm}

On the other hand, like the serial decoding of classical CCs, the decoding of $Z$-errors in AQCTPCs can also be done serially, i.e., an inner decoding followed by an outer decoding. However, the decoding algorithm for classical CCs can not be used to decode $Z$-errors directly. Instead, a modified version of syndrome-based decoding is needed, as explained next.

Before performing the decoding, the ancilla needs to be measured first to determine the syndrome information. Denote   the encoded quantum basis states of the AQCTPC
   $ \mathcal{Q}$  by
\begin{equation}
\label{AQC encoding}
|u +\mathcal{C}_{C}^\bot\rangle   \equiv   \frac{1}{\sqrt{|\mathcal{C}_{C}^\bot|}}\sum_{v \in \mathcal{C}_{C}^\bot}|u +v \rangle,
\end{equation}
where $u \in \mathcal{C}_T$. Suppose that a $Z$-error $e_Z$ happens to the encoded sate  (\ref{AQC encoding}), then
\begin{equation}
\label{AQC encoding_with_errors}
|u +\mathcal{C}_{C}^\bot\rangle    \rightarrow   \frac{1}{\sqrt{|\mathcal{C}_{C} |}}\sum_{w \in \mathcal{C}_{C} }(-1)^{wu^T}|w+e_Z \rangle.
\end{equation}
 First we take the syndrome measurement using the inner parity check matrix $H_{c_1}$ to get the inner syndrome information
 \begin{equation}
 \label{InnerZsyn}
\Psi_{\textrm{I}_i}\equiv H_{c_1}(w_i+e_{Z_i})^T =H_{c_1} e_{Z_i}^T,1\leq i\leq n_2,
\end{equation}
where  $w=(w_1,\ldots,w_{n_2})$ and $e_Z=(e_{Z_1},\ldots,e_{Z_{n_2}})$. The inner decodings are done first according to the inner syndrome information $ \Psi_{\textrm{I}_i}(1\leq i\leq n_2)$. They result in $n_2 $ decoded error sequences $\overline{e}_{Z_i}(1\leq i\leq n_2)$, each of length $n_1$. Denote by $\overline{e}_{Z}=(\overline{e}_{Z_1},\ldots,\overline{e}_{Z_{n_2}})$. We add the decoded result $\overline{e}_{Z}$ to  (\ref{AQC encoding_with_errors}), and then  perform the  measurement using the parity check matrix  $[H_{c_3}]\otimes  (I_{k_1},\textbf{0})$ to get the outer syndrome information
\begin{eqnarray}
\label{Zsynd22}
\nonumber
\Psi_{\textrm{O}}&\equiv&[H_{c_3}]\otimes  (I_{k_1},\textbf{0})(w +e_{Z}+\overline{e}_{Z})^T \\
&=&  [H_{c_3}]\otimes  (I_{k_1},\textbf{0})(e_{Z}+\overline{e}_{Z})^T.
\end{eqnarray}
 Discarding the zero part in $\Psi_{\textrm{O}}$ due to the $\textbf{0}$  sub-block   in $(I_{k_1},\textbf{0})$, the punctured $\Psi_{\textrm{O}}$ is then mapped into a sequence $\overline{\Psi}_{\textrm{O}}$  with elements over field $\mathbb{F}_{q^{k_1}}$.

The outer decoding is done with a syndrome-based decoding of $C_3$ according to the outer syndrome information $\overline{\Psi}_{\textrm{O}}$.  If the outer decoding is successful, we can obtain a decoded sequence $\mathbf{e}'_Z=(\mathbf{e}'_{Z_1},\ldots,\mathbf{e}'_{Z_{n_2}})$ with elements over $\mathbb{F}_{q^{k_1}}$. Then we map the sequence $\mathbf{e}'_Z$ back to the basis field $\mathbb{F}_q$, and  derive a   decoded error sequence $e'_Z=(e'_{Z_1},\ldots,e'_{Z_{n_2}})$ with  elements over $\mathbb{F}_{q }$, where $e'_{Z_i}(1\leq i\leq n_2)$ is the sub-sequence of length $k_1$. But notice that  $e'_Z$ is incomplete due to the $\textbf{0}$  sub-block in $(I_{k_1},\textbf{0})$.
In order to derive the fully-decoded error sequence, we only need to do some operations according  to the   inner syndrome information in (\ref{InnerZsyn}). Denote by $\widetilde{e}_Z=(\widetilde{e}_{Z_1},\ldots,\widetilde{e}_{Z_{n_2}})$ and $\widetilde{e}_{Z_i}=(e'_{Z_i}, f_{Z_i})(1\leq i\leq n_2)$, where $f_{Z_i}$ denotes the unknown errors   in $ \widetilde{e}_{Z_i}$ and is of length $r_1$.  Suppose that $H_{c_1}=(P_1,P_2)$, where $P_1$ is of size $r_1\times k_1$ and $P_2$ is an invertible $r_1\times r_1$ matrix, then we have $\Psi_{\textrm{I}_i}=H_{c_1}\widetilde{e}_{Z_i}^T=P_1e'^T_{Z_i} +P_2f_{Z_i}^T$   and then  $f_{Z_i}=[P_2^{-1}(\Psi_{\textrm{I}_i}-P_1e'^T_{Z_i})]^T$,   where $1\leq i\leq n_2$.

Similar to classical CCs, no matter how many $Z$-errors happen in each sub-block of length $n_1$, the outer decoding will not be affected provided that the total number of erroneous sub-blocks does not exceed the error correction ability of the outer code $C_3$. A summary of the full decoding process is provided in Algorithm \ref{DecCCs}.  A complexity analysis of the whole  decoding process follows.

In terms of decoding   $X$-errors with the TPC, we first need to   map the outer decoding  sequence from $\mathbb{F}_{q^{k_1}}$ to $\mathbb{F}_{q}$ whose running time complexity is $O(n_2)$ (see Algorithm \ref{DecTPCs}, lines  $11$-$15$). And it is easy to see that the complexity of the inner syndrome decoding of $C_1^\bot=[n_1,r_1]$  is $ O(1)$ since we just need to do $\widetilde{e}_{X_l}=(\Phi_{\textrm{I}_l},\textbf{0}) $, for $1\leq l\leq n_2$.  Therefore, the  inner decoding   complexity (IDC) of the TPC  is $O(n_2)$.  Recall that the IDC of  classical TPCs is  $\Omega(n_2a^{n_1})(a>1)$ by using the maximum likelihood (ML) decoding in general, which  is enormous   if $n_1$ is large.  Even though the inner codes can be efficiently decoded, see,  e.g., \cite{alhussien2010iteratively,wolf2006introduction}, the IDC of TPCs is still $\Omega(n_2n_1^b )(b>0)$.  Here, in quantum cases, we consider error degeneracy in the inner decoding and significantly reduce the IDC of TPCs to $O(n_2)$ in general.

 It is easy to see that the outer decoding complexity (ODC) of TPCs is completely determined by the outer constituent codes. Thus if  the outer codes  can be  decoded efficiently, the whole decoding of TPCs  is  efficient. For example, we let the outer codes be the Reed-Solomon (RS) codes or the generalized Reed-Solomon (GRS) codes that satisfy the dual-containing relationship \cite{ezerman2013pure,grassl2004optimal}. They  can be  decoded efficiently in time polynomial to their block length, e.g., by using the Berlekamp-Massey (BM) algorithm,
 see \cite{macwilliams1981theory,lin2004error,alhussien2010iteratively}. Then the whole decoding of TPCs for correcting   $X$-errors can be done  efficiently in polynomial time.
\begin{algorithm}[H]
\caption{The Decoding Algorithm of AQCTPCs for Correcting $Z $-errors.}
\label{DecCCs}
\begin{algorithmic}[1]
\Require
   $\Psi_{\textrm{I}_i}(1\leq i\leq n_2)$, $\Psi_{\textrm{O}}$, $H_{c_1}=(P_1,P_2)$, $H_{c_3}$;
\Ensure The decoded $Z $-error sequence $\widetilde{e}_Z$.
\State            Initialization: $ \overline{e}_{Z} = \emptyset,\widetilde{e}_Z=\emptyset$;
\For{ $i\in [1,n_2]$}
\State$//$ \verb"Do the inner decoding according to" (\ref{InnerZsyn}).
\State $H_{c_1} \overline{e}_{Z_i}^T=\Psi_{\textrm{I}_i}$, $\overline{e}_{Z}=(\overline{e}_{Z},\overline{e}_{Z_i}) $;
\EndFor
\State Map $\Psi_{\textrm{O}}$ to a sequence $\overline{\Psi}_{\textrm{O}}$ with elements over field $\mathbb{F}_{q^{k_1}}$;
\State $//$\verb"Do the outer decoding according to" $\overline{\Psi}_{\textrm{O}}$ \verb"and"  $C_3$.
\State$\mathbf{e}'_Z=(\mathbf{e}'_{Z_1},\ldots,\mathbf{e}'_{Z_{n_2}})$;
\For{  $i\in [1,n_2]$}
\State Map  $\mathbf{e}'_{Z_1}$ into a sequence over field   $\mathbb{F}_{q}$,  $e'_{Z_i}$;
\State $f_{Z_i}=[P_2^{-1}(\Psi_{\textrm{I}_i}-P_1e'^T_{Z_i})]^T$;
\State $\widetilde{e}_{Z_i}=(e'_{Z_i}, f_{Z_i})$;
\State $\widetilde{e}_{Z}=(\widetilde{e}_{Z},\widetilde{e}_{Z_i})$;
\EndFor \\
\Return $\widetilde{e}_{Z}$;
\end{algorithmic}
\end{algorithm}

 When correcting   $Z$-errors by using the CC, it is easy to see that the decoding complexity is the sum of the complexities of the inner and outer decodings. Thus, the CC is efficiently decodable  provided that the constituent codes $C_1$ and $C_3$ can be decoded efficiently, e.g., in time polynomial to the block length  \cite{macwilliams1981theory,lin2004error}. Overall, we can conclude that the entire AQCTPC decoding process for correcting both  $X$-errors and $Z$-errors is efficient provided that the inner and outer constituent codes are efficiently decodable.

  Similar to the generalization of classical CCs and TPCs, we can generalize the concatenation scheme of AQCTPCs by combining  GCCs with GTPCs.
 Let $A_\ell=[n_A,k_{\ell},d_{\ell}]_{q} (1\leq\ell\leq L)$ be $L$ $q$-ary linear codes. Let     $B_\ell=[N_B,K_{\ell},D_{\ell}]_{q^{k_{\ell}}} $ and $C_\ell=[N_B,M_{\ell},E_{\ell}]_{q^{k_{\ell}}}(1\leq\ell\leq L)$ be $L$ $q^{k_{\ell}}$-ary linear codes, respectively. Denote    $\mathbf{A}_\ell=[n_A,k_{\ell}^A]_{q} (1\leq\ell\leq L)$  by $L$   linear codes obtained by partitioning the generator matrix of $A_1$ into $L$ submatrices. Then we have the following result about the dual-containing relationship between GCCs and GTPCs.
\begin{lemma}
    Let
$
\mathcal{C}_{\mathcal{T}}
$
be the GTPC  of $\mathbf{A}_\ell^\bot$ and $B_\ell (1\leq\ell\leq L)$, and let
$\mathcal{C}_{\mathcal{C}}
$ be the GCC of $A_\ell $ and $C_\ell(1\leq\ell\leq L)$. If $ B_\ell^\bot\subseteq C_\ell$ for all $1\leq\ell\leq L$, then there is $\mathcal{C}_{\mathcal{T}}^\bot\subseteq \mathcal{C}_{\mathcal{C}}$.
\end{lemma}
\begin{IEEEproof}
We use  the notations for   GTPCs and GCCs given in    Preliminaries.
Denote  the \textit{collection of duals matrix} (cdm) (see Ref. \cite{maucher2000equivalence}) of $G_1^A$  in (\ref{G1A}) by
\begin{equation}
 \hat{H}^A=\textrm{cdm}(G^A_1)=\left(
\begin{array}{c}
 \hat{H}^A_1\\

\hat{H}^A_2\\
\vdots\\

\hat{H}^A_{L+1}
\end{array}
\right)
\end{equation}
 with $k_\ell^A=\textrm{rank}(\widehat{H}^A_\ell)=\textrm{rank}(\mathbf{G}_\ell^A) $, for $1\leq\ell\leq L$, and $k_{L+1}^A=\textrm{rank}(\widehat{H}^A_{L+1})=n_A-k_1$.
 Then the parity check matrix of the GCC $\mathcal{C}_\mathcal{C}$ is given by
\begin{equation}
 H_{\mathcal{C}_\mathcal{C}}\equiv\left(
\begin{array}{l}
[H^{C}_1]\otimes \widehat{H}^A_1\\

\hspace{9mm}\vdots\\

[H^{C}_L]\otimes \widehat{H}^A_{L}\\

[I_{N_B}]\otimes \widehat{H}^A_{L+1}
\end{array}
\right).
\end{equation}
And the parity check matrix of the GTPC $\mathcal{C}_\mathcal{T}$ is given by
\begin{equation}
 H_{\mathcal{C}_{\mathcal{T}}}\equiv\left(
\begin{array}{c}
[H^{B^t}_1]\otimes \mathbf{G}^A_1\\

[H^{B^t}_2]\otimes \mathbf{G}^A_2\\
\vdots\\

[H^{B^t}_L]\otimes \mathbf{G}^A_L
\end{array}
\right)
\end{equation}
According to Ref. \cite{maucher2000equivalence} and Ref. \cite{fan2017comments}, we know   the following two properties about the cdm of $G_1^A$:
 \begin{itemize}
 \item  $\hat{H}_\ell^A\mathbf{G}^{A^t}_\hbar=0$, for all $1\leq \ell\leq  L +1$, $1\leq \hbar\leq  {L}$ and $\ell\neq \hbar$.
 \item  $\hat{H}_\ell^A\mathbf{G}^{A^t}_\ell $ is of full rank, for all $1\leq \ell\leq L $.
 \end{itemize}
Since $\hat{H}_\ell^A\mathbf{G}^{A^t}_\ell $ is of full rank, we can always find an invertible matrix $U_\ell$ such that $U_\ell\hat{H}_\ell^A\mathbf{G}^{A^t}_\ell $ is an identity matrix, for $1\leq \ell\leq L $. If  $ B_\ell^\bot\subseteq C_\ell$,   which means that $[H_\ell^C][H_\ell^{B^t}]^T=0$ for all $1\leq\ell\leq L$, then there is $H_{\mathcal{C}_\mathcal{C}}H_{\mathcal{C}_\mathcal{T}}^T=0$ and we have $\mathcal{C}_{\mathcal{T}}^\bot\subseteq \mathcal{C}_{\mathcal{C}}$.
\end{IEEEproof}
\begin{theorem}
\label{GeneralizedAQCTPCTheorem}
   There exist  generalized AQCTPCs  with parameters \begin{equation}
   \mathcal{Q} =[[N_Bn_A,\sum_{\ell=1}^L(K_\ell+M_\ell-N_B)k_\ell^A,d_Z/d_X]]_q,
\end{equation}
   where $d_Z\geq\min\{D_1d_1 ,\ldots,D_Ld_L\}$, $d_X\geq\min\{E_1 ,\ldots, E_L \}$.
\end{theorem}
\begin{IEEEproof}
By combining   Lemma \ref{AQC Constructions} and Lemma \ref{asymmetric tensor and concatenated}, we can obtain the generalized AQCTPCs with parameters
\begin{equation}
   \mathcal{Q} =[[N_Bn_A,\sum_{\ell=1}^L(K_\ell+M_\ell-N_B)k_\ell^A,d_Z/d_X]]_q.
 \end{equation}
We use the GCCs to correct $Z$-errors and thus  the $Z$-distance  $d_Z$ of the generalized AQCTPC  $\mathcal{Q}$ is given by $d_Z\geq\min\{D_1d_1 ,\ldots,D_Ld_L\}$. Next we need to compute the  $X$-distance  $d_X$ of $\mathcal{Q}$.  Suppose that there is an   $X$-error $e_X$ of length $N_Bn_A$   in the encoded codeword.  Denote
\begin{eqnarray}
 \nonumber
 \Phi_X&\equiv &H_{\mathcal{C}_\mathcal{T}} e_X^T \\
&=&  \left(
\begin{array}{c}
[H^{B^t}_1]\otimes \mathbf{G}^A_1\cdot e_X^T \\

[H^{B^t}_2]\otimes \mathbf{G}^A_2\cdot e_X^T\\
\vdots\\

[H^{B^t}_L]\otimes \mathbf{G}^A_L\cdot e_X^T
\end{array}
\right)
\end{eqnarray}
by the syndrome information obtained by measuring the ancilla and let $\Phi_{\textrm{X}_\ell}\equiv[H^{B^t}_1]\otimes \mathbf{G}^A_\ell\cdot e_X^T $, $1\leq\ell\leq L$. Suppose that for some $1\leq\imath\leq L$, we have $E_\imath =\min\{E_1 ,\ldots,E_L \}$. Similar to the proof of Theorem  \ref{AQCTPCTheorem}, if  $ \textrm{wt}(e_X)\leq E_\imath -1$, then we must have $\Phi_{\textrm{X}_\imath} \neq 0$ and then the error  can be detected or  $\Phi_{\textrm{X}_\imath} = 0$ but the error is degenerate. Therefore we have $d_X\geq\min\{E_1 ,\ldots, E_L \}$.
\end{IEEEproof}

It should be noticed  in the proof of Theorem \ref{GeneralizedAQCTPCTheorem} that, we only give  a minimum limit of the distance $d_X$. In the practical error correction, e.g., in \cite{spinner2016soft} for classical GCCs, we have $L$ syndrome information $\Phi_{\textrm{X}_\ell}(1\leq\ell\leq L)$ to be used for the decoding and then the generalized AQCTPCs can correct many more $X$-errors beyond the minimum distance limit in Theorem \ref{GeneralizedAQCTPCTheorem} in practice.

\section{Families of AQCTPCs}
 \label{Families of AQCTPCs}
 In this section, we provide examples of AQCTPCs that outperform best-known AQCs in the literature.
 Since the inner constituent codes  $C_1$ in AQCTPCs can be chosen arbitrarily, we can get varieties of  AQCTPCs  by using different types of the constituent codes. Although the construction of  AQCTPCs is not restricted by the field size $q$,  in this section, we mainly focus on binary codes which may be more practical in the future application. For simplicity,  if $q=2$, we omit the subscript  in the parameters of quantum and classical codes.

Firstly we use classical single-parity-check codes \cite{macwilliams1981theory} as the inner constituent codes and we have the following result.
\begin{corollary}
\label{corollary_reptition}
There exists  a family of binary AQCTPCs with parameters
\begin{equation}
\label{AQCTPCeven}
\mathcal{Q}=[[N_Q,K_\mathcal{Q},d_Z\geq2d_3/d_X\geq d_2]],
\end{equation}
\end{corollary}
where $N_Q=(m_1+1)n_2$, $K_\mathcal{Q}= m_1(n_2-d_2-d_3+2)$, $m_1\geq2$, $2\leq n_2\leq 2^{m_1}+1$, and $2\leq d_2+d_3 \leq n_2 +2$ are all integers.
\begin{IEEEproof}  Let $C_1=[m_1+1,m_1,2]$  be a binary  single-parity-check code with even codewords, and let $C_2=[n_2,k_2,d_2]_{2^{m_1}}$  and $C_3=[n_2,k_3,d_3]_{2^{m_1}}$ be two classical GRS codes.
It is shown in \cite{ezerman2013pure} that if  $2\leq n_2\leq 2^{m_1}+1$ and $2\leq d_2+d_3 \leq n_2 +2$,  there exists    $C_3^\bot\subseteq C_2$.
\end{IEEEproof}

In Corollary \ref{corollary_reptition}, if we let $d_2=2d_3$, then we can also obtain  a family  of   symmetric quantum codes with parameters \begin{equation}\label{AQCTPCSymmetry}
\mathcal{Q}=[[N_\mathcal{Q},K_\mathcal{Q},d_\mathcal{Q}\geq d_2]],
 \end{equation}
 where $N_\mathcal{Q}=(m_1+1)n_2$, $K_\mathcal{Q}=m_1(n_2-3d_2/2+2)$, $2\leq d_2\leq 2(n_2+2)/3 $. We first compare (\ref{AQCTPCSymmetry}) with QBCH codes in \cite{aly2007quantum}. It is known that the minimum distance of QBCH codes of length $\Theta(N_\mathcal{Q})$ is upper bounded by $ c\sqrt{N_\mathcal{Q}}$  ($c>0$ is a constant).  On the other hand, the minimum distance of our codes is upper bounded by $2(n_2+2)/3$ which is   larger than $ c\sqrt{N_\mathcal{Q}}$ provided that $n_2\geq9c^2(m_1+1)/4-4 $.

  For example,  let $n_2=2^{m_1}+1$, then the minimum distance of our codes  can be as large as $2N_\mathcal{Q}/(3\log(N_\mathcal{Q}))$, which is almost linear to the length $N_\mathcal{Q}$, while the dimension is larger than $\log(N_\mathcal{Q})$.    If we let $
  d_2=O((N_\mathcal{Q})^{c_1}),
$
 where $1/2<c_1<1$ is any  constant, then the  rate  of our codes
 \begin{equation}
 R_\mathcal{Q}=\frac{K_\mathcal{Q}}{N_\mathcal{Q}}=\frac{m_1}{m_1+1}(1-\frac{3d_2}{2n_2}+\frac{2}{n_2})
 \end{equation}
   is equal to   $1$  as $n_2=2^{m_1}+1$ goes to infinity and $d_\mathcal{Q}\geq d_2=O((N_\mathcal{Q})^{c_1})$.   In \cite{aly2007quantum}, the rate of binary QBCH codes of CSS type   is given by
   \begin{equation}
 R_{QBCH}=1-\frac{m(\delta-1)}{N},
 \end{equation}
 where $N$ is the block length, $m=\textrm{ord}_N(2)$  is the multiplicative order of $2$ modulo $N$, and $2\leq\delta\leq N(2^{\lceil m/2\rceil}-1)/(2^m-1)=O(\sqrt{N})$. It is easy to see that $R_{QBCH}$  is also asymptotic to $1$ as $N$ goes to infinity. However
our codes have much better minimum distance upper bound than QBCH codes.

\begin{table*}
\caption{Comparison of binary AQCTPCs with the binary extension of asymmetric QRS codes in \cite{la2012asymmetric}. The AQCTPCs are derived from Corollary \ref{corollary_reptition}.
The $``-"$ in the table means that there do not exist AQCs with comparable  parameters in Ref. \cite{la2012asymmetric}.  In quantum codes, an AQC with parameters $[[n,0,d_Z/d_X]] $  of dimension $1$ is a pure   state which can correct all $X$-errors of weight up to $\lfloor(d_X-1)/2\rfloor$  and all $Z$-errors of weight up to $\lfloor(d_Z-1)/2\rfloor$ \cite{calderbank1998quantum,frederic2013csslike}.  To
facilitate notation,    the numbers of $Z$- and $X$-distance of the  AQCs   are the  lower bound. }
\setlength{\tabcolsep}{3.2pt}
\centering
\begin{tabular}[c]{|c|l|l|c|l|l|c|l|l|l|}
\hline
$m_1$&AQCTPCs  &Ref. \cite{la2012asymmetric} &$m_1$ & AQCTPCs&Ref. \cite{la2012asymmetric} &$m_1$ & AQCTPCs&Ref. \cite{la2012asymmetric}\\
\hline
$6$&$[[378,6,104/3]]$&$-$&$7$&$[[888,7,218/3]]$&$-$&$8$&$[[2034,8,448/3]]$&$-$\\
 \hline
$6$&$[[378,12,102/3]]$&$-$&$7$&$[[888,14,216/3]]$&$-$&$8$&$[[2034,16,446/3]]$&$-$\\
 \hline
$6$&$[[378,132,62/3]]$&$[[378,0 ,62/3]]$&$7$ &$[[888,329,126/3]]$&$[[889,0,126/3]]$&$8$&$[[2034,784,254/3]]$&$[[2040,0,254/3]]$\\
 \hline
$6$&$[[378,138,60/3]]$&$[[378,12,60/3]]$&$7$ &$[[888,336,124/3]]$&$[[889,14,124/3]]$&$8$&$[[2034,792,252/3]]$&$[[2040,16,252/3]]$\\
 \hline
\multicolumn{1}{|c|}{$\vdots$}&\multicolumn{1}{c|}{$\vdots$}&\multicolumn{1}{c|}{$\vdots$}&\multicolumn{1}{c|}{$\vdots$}&\multicolumn{1}{c|}{$\vdots$}&\multicolumn{1}{c|}{$\vdots$}&\multicolumn{1}{c|}{$\vdots$}&\multicolumn{1}{c|}{$\vdots$}&\multicolumn{1}{c|}{$\vdots$}\\
 \hline
$6$&$[[378,258,20/3]]$&$[[378,252,20/3]]$&$7$ &$[[888,651,34/3]]$&$[[889,644,34/3]]$&$8$&$[[2034,1560,60/3]]$&$[[2040,1552,60/3]]$\\
 \hline
$6$&$[[378,6,100/5]]$&$-$&$7$ &$[[888,7,214/5]]$&$-$&$8$&$[[2034,8,444/5]]$&$-$\\
 \hline
$6$&$[[378,12,98/5]]$&$-$&$7$ &$[[888,14,212/5]]$&$-$&$8$&$[[2034,16,442/5]]$&$-$\\
 \hline
$6$&$[[378,126,60/5]]$&$[[378,0,60/5]]$&$7$ &$[[888,322,124/5]]$&$[[889,0,124/5]]$&$8$&$[[2034,776,252/5]]$&$[[2040,0,252/5]]$\\
 \hline
$6$&$[[378,132,58/5]]$&$[[378,12,58/5]]$&$7$ &$[[888,329,122/5]]$&$[[889,14,122/5]]$&$8$&$[[2034,784,250/5]]$&$[[2040,16,250/5]]$\\
 \hline
\multicolumn{1}{|c|}{$\vdots$}&\multicolumn{1}{c|}{$\vdots$}&\multicolumn{1}{c|}{$\vdots$}&\multicolumn{1}{|c|}{$\vdots$} &\multicolumn{1}{c|}{$\vdots$}&\multicolumn{1}{c|}{$\vdots$}&\multicolumn{1}{c|}{$\vdots$}&\multicolumn{1}{c|}{$\vdots$}&\multicolumn{1}{c|}{$\vdots$}\\
 \hline
$6$&$[[378,246,20/5]]$&$[[378,240,20/5]]$&$7$ &$[[888,637,34/5]]$&$[[889,630,34/5]]$&$8$&$[[2034,1544,60/5]]$&$[[2040,1536,60/5]]$\\
\hline
\end{tabular}\label{ComparisonsMDS1}
\end{table*}
 Then we compare AQCTPCs in Corollary \ref{corollary_reptition} with the extension of asymmetric quantum  MDS codes in \cite{la2012asymmetric}. For simplicity, we consider the extension of  binary asymmetric QRS codes   in \cite{la2012asymmetric} with parameters
\begin{equation}
\label{AQCs_RS}
 [[ \widetilde{N}_\mathcal{Q} , \widetilde{K}_\mathcal{Q} , \widetilde{d}_Z\geq \widetilde{d}_1/ \widetilde{d}_X\geq \widetilde{d}_2]] ,
\end{equation}
where $\widetilde{N}_\mathcal{Q}=m_1(2^{m_1}-1)$, $\widetilde{K}_\mathcal{Q}={m_1}(2^{m_1}-\widetilde{d}_1-\widetilde{d}_2+1)$, $2\leq \widetilde{d}_1+\widetilde{d}_2\leq2^{m_1}+1$. In order to make a fair
comparison between them, we let   $ n_2=\lfloor m_1(2^{m_1}-1)/(m_1+1)\rfloor$ in Corollary \ref{corollary_reptition} so that they have    an equal or a similar block length. Let $\widetilde{d}_1=2d_3$ and $\widetilde{d}_2=d_2$, then it is easy to see that if $d_3\geq  (2^{m_1}-1)/(m_1+1)$, the dimension of AQCTPCs in (\ref{AQCTPCeven}) is larger than that of AQCs in (\ref{AQCs_RS}). Further, AQCTPCs of length $N_Q=\Theta(m_1(2^{m_1}-1))$  in Corollary \ref{corollary_reptition}   can be decoded efficiently in polynomial time and we have the following result.
\begin{corollary}
\label{AQCTPCsComplexityAna}
There exist AQCTPCs of length $N_Q=\Theta(m_1(2^{m_1}-1))$ which can be decoded in $O(N_Q^2/\log N_Q)$  arithmetic operations.
\end{corollary}
\begin{IEEEproof}
First we consider the  complexity of the decoding of $X$-errors. The IDC  of TPCs is $O(2^{m_1})$ according to the proof of Theorem \ref{AQCTPCTheorem}. It is known that  GRS codes of length $n_2=\Theta(2^{m_1})$ can be decoded in  $O(n_2^2)$ field operations by using the BM algorithm \cite{macwilliams1981theory,sarwate1977complexity}. Therefore the total decoding   of TPCs requires  $O(n_2^2)$ arithmetic operations.

Then we consider the decoding of $Z$-errors by using  the CC. If we use Algorithm \ref{DecCCs} to do the decoding, we can only decode up to $\lfloor (2d_3-1)/4\rfloor$ numbers of $Z$-errors. In order to decode any $Z$-error  of weight smaller than  half   the minimum distance  $2d_3$, we  use the inner code $C_1=[m_1+1,m_1,2]$ to do the error detection for each sub-block of the CC. Suppose we can detect   $t_1$ erroneous sub-blocks  and suppose that  there exist $t_2$ erroneous sub-blocks  which are undetectable. As a result, there are $t_1$ erroneous positions which are known in the error sequence  $\mathbf{e}_Z$ that corresponds  to the outer code and, $ t_2$ erroneous positions that are unknown. It is easy to see that if the weight of the $Z$-error is smaller than $d_3$,   we must have $0\leq t_1+2t_2\leq2d_3-1$. Then we can decode the error sequence $\mathbf{e}_Z$ with $t_1$   errors  in known locations and $t_2$ errors which are unknown by using the BM algorithm in $O(n_2^2)$ arithmetic operations \cite{macwilliams1981theory,forney1966generalized}.   Therefore the total complexity of decoding $Z$-errors is $O(n_2^2)$. Overall, the whole decoding of AQCTPCs of length $N_Q=\Theta(m_1(2^{m_1}-1))$ can be fulfilled in $O(N_Q^2/\log N_Q)$  arithmetic operations.
\end{IEEEproof}

In
Table~\ref{ComparisonsMDS1}, we make a   comparison between parameters of AQCTPCs and the results in \cite{la2012asymmetric}. It is shown that   AQCTPCs   have a relatively large $Z$-distance and can have much better parameters than the codes in \cite{la2012asymmetric}. Further, the decoding complexity of AQCs  in \cite{la2012asymmetric} of length $\widetilde{N}_Q= m_1(2^{m_1}-1) $ is dominated by the RS code decoding whose complexity   scales as $O(\widetilde{n}_2^2)$ by using the BM decoder, where $\widetilde{n}_2=2^{m_1}-1$. Then the decoding complexity of AQCs in \cite{la2012asymmetric} is $O(\widetilde{N}_Q^2/\log \widetilde{N}_Q)$. It is shown that both   AQCTPCs in Corollary  \ref{AQCTPCsComplexityAna} and AQCs in \cite{la2012asymmetric}   can  be decoded efficiently in polynomial arithmetic operations. However,  our codes have much better parameters than the  codes in \cite{la2012asymmetric}.

Next we use classical Simplex codes \cite{macwilliams1981theory} that have a   large minimum distance as the inner constituent codes. We show that Simplex codes   can  result in  AQCTPCs with a large $Z$-distance  $d_Z$. \begin{corollary}
\label{corollary_simplex}
There exists  a family of binary AQCTPCs with parameters
\begin{equation}
\mathcal{Q}=[[N_\mathcal{Q} ,K_\mathcal{Q} ,d_Z\geq 2^{m_1-1}d_3/d_X\geq d_2]],
\end{equation}
where $N_\mathcal{Q}=(2^{m_1}-1)n_2$, $K_\mathcal{Q}=m_1(n_2-d_2-d_3+2),$$m_1\geq 2$, $2\leq n_2\leq 2^{m_1}+1$, and $2\leq d_2+d_3 \leq n_2 +2$.
\end{corollary}
\begin{IEEEproof}
The proof proceeds in the same way as in Corollary~\ref{corollary_reptition} except we use  classical Simplex codes  $C_1=[2^{m_1}-1,m_1,2^{m_1-1}]$ as the inner constituent code.
\end{IEEEproof}

In particular, if we take $n_2=2^{m_1}+1$ and let    $d_2=O(2^{cm_1})$ and $d_3=2^{m_1}+2-d_2$, where $0<c<1$ is a constant, then we have
\begin{equation}
\mathcal{Q} =[[N_\mathcal{Q},m_1,  d_Z   /  d_X]],
\end{equation}
where $N_\mathcal{Q} =2^{2m_1}-1$, $ d_Z \geq 2^{m_1-1}(2^{m_1}+2-d_2)$, and $d_X\geq d_2$. It is easy to see that $d_Z/N_\mathcal{Q}\rightarrow1/2$ as $m_1\rightarrow \infty$ and $\mathcal{Q}$ can meet the quantum Gilbert-Varshamov (GV)  bound for AQCs in \cite{matsumoto2017two}. Therefore we get a family of AQCTPCs with a very large  $Z$-distance   $d_Z$ which is of approximately half   the block length, at the same time, the dimension and the  $X$-distance   $d_X$ can continue increasing as the block length goes to infinity. In Table \ref{ComparisonsSimplex}, we list several AQCTPCs with a large  $Z$-distance   $d_Z$ which is of approximately half   the block length.  In particular, if  $d_X=2$, then the $Z$-distance   $d_Z$ of AQCTPCs in Corollary \ref{corollary_simplex} could be larger than half   the block length.

\begin{table}[h!]
 \setlength{\tabcolsep}{3pt}
 \caption{Construction of binary AQCTPCs whose  $Z$-distance   $d_Z$ is approximately half of the block length by using binary Simplex codes   as   inner codes $C_1$.  The outer codes $C_2$ and $C_3$ are dual-containing MDS codes derived from \cite{ezerman2013pure}. To
facilitate notation,   the numbers of $Z$- and $X$-distance of the  AQCTPCs  are the  lower bound.  }
\centering
\begin{tabular}{|p{10pt}|p{10pt}|p{70pt}||p{10pt}|p{10pt}|p{75pt}|}
\hline
 $m_1$   &  $d_2 $  & AQCTPCs& $m_1$   &  $d_2 $  & AQCTPCs\\
\hline
$2$  &  $2$ &$ [[15, 2, 8/2]]$& $6$  &  $4$ &$ [[4095,6, 1984/4]]$\\ \hline
$2$  &  $3$ &$ [[15, 2, 6/3]]$&$6$  &  $5$ &$ [[4095,6, 1952/5]]$\\ \hline
$3$  &  $2$ &$ [[63, 3, 32/2]]$& $6$  &  $6$ &$ [[4095,6, 1920/6]]$\\ \hline
$4$  &  $2$ &$ [[255, 4, 128/2]]$& $6$  &  $7$ &$ [[4095,6, 1888/7]]$\\ \hline
$5$  &  $2$ &$ [[1023,5, 512/2]]$& $7$  &  $2$ &$ [[16383,7, 8192/2]]$ \\ \hline
$5$  &  $3$ &$ [[1023,5, 496/3]]$&  $7$  &  $3$ &$ [[16383,7, 8128/3]]$\\ \hline
$5$  &  $4$ &$ [[1023,5, 480/4]]$&$7$  &  $4$ &$ [[16383,7, 8064/4]]$ \\ \hline
$5$  &  $5$ &$ [[1023,5, 464/5]]$&$7$  &  $5$ &$ [[16383,7, 8000/5]]$ \\ \hline
$6$  &  $2$ &$ [[4095,6, 2048/2]]$&$7$  &  $6$ &$ [[16383,7, 7936/6]]$ \\ \hline
$6$  &  $3$ &$ [[4095,6, 2016/3]]$&$7$  &  $7$ &$ [[16383,7, 7827/7]]$ \\
\hline
\end{tabular}
\label{ComparisonsSimplex}
\end{table}

In addition, if we use linear codes  in \cite{Grassl:codetables} with best known parameters  as the inner codes, we can get many new  AQCTPCs with a relatively  large  $Z$-distance   $d_Z$ and  very flexible   code parameters. We list some of them in Table \ref{ComparisonsOnline}.  The $Z$-distances of  the last four codes in Table \ref{ComparisonsOnline}    are much larger than half   the block length, respectively.  All the AQCTPCs  in   Table \ref{ComparisonsSimplex} and Table \ref{ComparisonsOnline}    have the largest  $Z$-distance   $d_Z$ compared to  existed AQCs with comparable    block length and  $X$-distance  $d_X$.
\begin{table}[h!]
 \setlength{\tabcolsep}{3pt}
 \caption{ Construction of binary AQCTPCs with a large  $Z$-distance by using some best known linear codes in Ref. \cite{Grassl:codetables} as   inner codes. The outer codes $C_2=[n_2,k_2,d_2]_{2^{k_1}}$ and $C_3=[n_2,k_3,d_3]_{2^{k_1}}$ are dual-containing MDS codes with optimal parameters in \cite{ezerman2013pure}, respectively. To
facilitate notation,   the numbers of $Z$- and $X$-distance of the  AQCTPCs in this table are the  lower bound.  }
\centering
\begin{tabular}{|p{55pt}|p{55pt}|p{70pt}|}
\hline
 $C_1 $ in Ref. \cite{Grassl:codetables}   &   $\{n_2,d_2,d_3\} $   \par in Theorem \ref{AQCTPCTheorem}    & AQCTPCs \\
\hline
$[7,3,4] $  &  $\{9,3,7\}$ &$ [[63, 3, 28/3]]$ \\
\hline
$[7,3,4]$  & $\{9,5,5\}$&$ [[63, 3, 20/5]]$ \\
\hline
$[8,4,4]$  &  $\{17,3,15\}$ &$ [[136, 4, 60/3]]$ \\
\hline
$[8,4,4]$  & $\{17,5,13\}$&$ [[136, 4, 52/5]]$ \\
\hline
$[12,4,6]$  & $\{17,3,15\}$&$ [[204, 4, 90/3]]$ \\
\hline
$[12,4,6]$  & $\{17,5,13\}$&$ [[204, 4, 78/5]]$ \\
\hline
$[15,4,8]$  & $\{17,3,15\}$&$ [[255, 4, 120/3]]$\\
\hline
$[15,4,8]$  & $\{17,5,13\}$&$ [[255, 4, 104/5]]$\\
\hline
$[16,5,8]$  & $\{33,3,31\}$&$ [[528, 5, 248/3]]$\\
\hline
$[16,5,8]$  & $\{33,5,29\}$&$ [[528, 5, 232/5]]$\\
\hline
$[21,5,10]$  & $\{33,3,31\}$&$ [[693, 5, 310/3]]$\\
\hline
$[21,5,10]$  & $\{33,5,29\}$&$ [[693, 5, 290/5]]$\\
\hline
$[22,6,9]$  & $\{65,3,63\}$&$ [[1430, 6, 567/3]]$\\
\hline
$[22,6,9]$  & $\{65,5,61\}$&$ [[1430, 6, 549/5]]$\\
\hline
$[24,7,10]$  & $\{129,3,127\}$&$ [[3096, 7, 1270/3]]$\\
\hline
$[24,7,10]$  & $\{129,5,125\}$&$ [[3096, 7, 1250/5]]$\\
\hline
$[63,3,36]$  & $\{9,2,8\}$&$ [[567, 3, 288/2]]$\\
\hline
$[127,3,72]$  & $\{9,2,8\}$&$ [[1143, 3, 576/2]]$\\
\hline
$[255,3,145]$  & $\{9,2,8\}$&$ [[2295, 3, 1160/2]]$\\
\hline
$[255,4,136]$  & $\{17,2,16\}$&$ [[4335, 4, 2176/2]]$\\
\hline
\end{tabular}
\label{ComparisonsOnline}
\end{table}

If we use asymptotically good linear codes that can attain the classical GV bound as the inner codes $C_1$, we can get the following asymptotic result about AQCTPCs.
\begin{corollary}
\label{AQECC corollary}
There exists a family of $q$-ary AQCTPCs with parameters \begin{equation}\mathcal{Q} =[[N_\mathcal{Q}=n_1n_2,K_\mathcal{Q},  d_Z/  d_X]]_q \end{equation} such that
\begin{eqnarray}
\label{asymptoticGV}
  \frac{K_\mathcal{Q}}{N_\mathcal{Q}} &\geq&   \left(1-H_q\left(\frac{d_1}{ n_1}\right)\right)\left(1- \frac{d_2}{n_2}- \frac{d_3}{n_2} \right), \\
  d_Z&\geq& d_1d_3,\\
  d_X&\geq& d_2,
\end{eqnarray}
where
\begin{equation}
H_q(x)=x\log_q(q-1)-x\log_qx-(1-x)\log_q(1-x)
\end{equation}
 is the $q$-ary entropy function, $2\leq d_1\leq n_1$, $2\leq d_2+d_3\leq n_2$, and $n_1, n_2\rightarrow  \infty$.
\end{corollary}
\begin{IEEEproof}
We choose $C_1=[n_1,k_1,d_1]_q$ to be asymptotically good linear codes meeting the GV bound, i.e.,
\begin{equation}
\frac{k_1}{n_1}\geq 1-H_q(\frac{d_1}{n_1}).
 \end{equation}
 Let $C_2=[n_2,k_2,d_2]_{q^{k_1}}$ and $C_3=[n_2,k_3,d_3]_{q^{k_1}}$ be two MDS  codes such that $C_2^\bot\subseteq C_3$. Denote by $N_\mathcal{Q}=n_1n_2$, $K_\mathcal{Q}=k_1(k_2+k_3-n_2)$, $d_Z=d_1d_3$  and $d_X=d_2$. According to Theorem \ref{AQCTPCTheorem}, we can get the asymptotic result in (\ref{asymptoticGV}) as $n_1,n_2\rightarrow  \infty$.
\end{IEEEproof}

On the other hand, besides using   MDS codes as the outer constituent codes, we can also use AG codes that satisfy the dual-containing constraint \cite{ashikhmin2001asymptotically,wang2010asymmetric}.  We will adopt the notation of AG codes used in \cite{tsfasman1991algebraic,wang2010asymmetric}.
\begin{theorem}[\cite{wang2010asymmetric}]
\label{quantumAGwang}
Let $\mathcal{X}$ be an algebraic curve  over $\mathbb{F}_q$ of genus $g$ with at
least $n$  rational points. For any $2g-2<s<l<n$, there exist two $q$-ary AG codes $C_1=[n,k_1,d_1]_q$ and  $C_2=[n,k_2,d_2]_q$ with $k_1=n-k_2+l-s$ such that $C_2^\bot\subset C_1$,   where   $d_1\geq s-2g+2$ and $d_2\geq n-l$.
\end{theorem}

For $q=2^m(m\geq2)$,  there is the following asymptotic result about asymmetric QAG codes in \cite{wang2010asymmetric}.
\begin{theorem}[\cite{wang2010asymmetric}]
\label{AsymptoticAGwang}
Let $ q=2^m$ and let $0\leq\delta_x,\delta_z\leq1$ such that $\delta_x+\delta_z\leq 1-2/(\sqrt{2^{m}}-1)$, then there
exists a family of asymptotically good asymmetric QAG codes $\mathcal{Q}$ satisfying
\begin{equation}
R_\mathcal{Q}(\delta_x,\delta_z)\geq1-\delta_x-\delta_z-\frac{2}{\sqrt{2^{m}}-1}.
\end{equation}
\end{theorem}

By using similar code extension  methods in \cite{ashikhmin2001asymptotically} and the CSS construction of AQCs, one can   obtain asymptotically good   binary extensions of asymmetric QAG codes as follows.
\begin{corollary}
\label{AsymmetricAGwang}
Let $ q=2^m$ and let $0\leq\delta_x,\delta_z\leq1$ such that $\delta_x+\delta_z\leq 1-2/(\sqrt{2^{m}}-1)$, then there
exists a family of  asymptotically good binary asymmetric QAG codes $\mathcal{Q}$ satisfying
\begin{equation}
\label{asymptoticQAG}
R_\mathcal{Q}(\delta_x,\delta_z)\geq1-m\delta_x-m\delta_z-\frac{2}{\sqrt{2^{m}}-1}.
\end{equation}
\end{corollary}
\begin{IEEEproof}
The asymptotic bound in (\ref{asymptoticQAG}) can be obtained from Ref.~\cite{ashikhmin2001asymptotically} and Theorem \ref{AsymptoticAGwang}.
\end{IEEEproof}

 Denote by $C_1=[n_1,k_1,d_1]$  a  binary linear code and let $\mathcal{X}$ be an algebraic curve  over $\mathbb{F}_{2^{k_1}}$ of genus $g$ with at
least $n_2$  rational points. Then we have the following result for constructing AQCTPCs by using AG codes as   outer codes.
\begin{proposition}
\label{asymmetricquantumAGnew}
 There exists a family of binary AQCTPCs with   parameters
\begin{equation}
  \label{AQCTPCAGs}
\mathcal{Q}=[[N_\mathcal{Q} ,K_\mathcal{Q} ,d_Z\geq d_1d_3/d_X\geq d_2]],
\end{equation}
 where $N_\mathcal{Q}=n_1n_2$, $K_\mathcal{Q}=k_1(l-s)$, $2g-2<s<l<n_2$, $d_2\geq s-2g+2$ and $d_3\geq n_2-l$. As $n_2$ goes to infinity, the following asymptotic bound of AQCTPCs holds
  \begin{equation}
  \label{asymptoticAQCTPCs}
 R_{\mathcal{Q}}\geq \frac{k_1}{n_1}\left(1-\frac{n_1}{d_1}\delta_z-n_1\delta_x-\frac{2}{\sqrt{2^{k_1}}-1}\right),
 \end{equation}
 where $R_\mathcal{Q}=K_\mathcal{Q}/N_\mathcal{Q}$, $ \delta_X$ and $\delta_Z$ are the relatively minimum distance of $\mathcal{Q}$.
\end{proposition}
\begin{IEEEproof}
According to Theorem \ref{quantumAGwang}, we know that there exist    two $2^{k_1}$-ary AG codes    $C_2=[n_2,k_2,d_2]_{2^{k_1}}$ and  $C_3=[n_2,k_3,d_3]_{2^{k_1}}$     such that $C_2^\bot\subseteq C_3$, where $k_2=n_2-k_3+l-s$ and $2g-2<s<l<n_2$. Then from  Theorem \ref{AQCTPCTheorem},  we can construct a family of binary AQCTPCs with parameters
$\mathcal{Q}=[[N_\mathcal{Q}=n_1n_2,K_\mathcal{Q}=k_1(l-s),d_Z\geq d_1d_3/d_X\geq d_2]]$, where $d_2\geq s-2g+2$ and $d_3\geq n_2-l$. Denote by $ \delta_X$ and $\delta_Z$  the relatively minimum distance of $\mathcal{Q}$, i.e., $\delta_X=d_X/N_\mathcal{Q}$ and $\delta_Z=d_Z/N_\mathcal{Q}$.  The asymptotic  result  can be obtained by  Theorem \ref{AsymptoticAGwang}.
\end{IEEEproof}

In Fig. $2$, we compare the asymptotic bound of AQCTPCs in (\ref{asymptoticAQCTPCs}) with that of asymmetric QAG codes in (\ref{asymptoticQAG}).  We also give the  GV bound of CSS codes for comparisons. In order to  get as   good as possible
 asymptotic curves for AQCTPCs, we  use different inner constituent codes to generate several piecewise asymptotic curves and then joint them  together. In  Fig. $2$(a), we can see that the asymptotic bound of AQCTPCs is better than
that for  asymmetric QAG codes when the relative minimum distance $0.02<\delta_Z<0.06$. As the  the asymmetry $\theta =d_Z/d_X $ grows, it is shown in Fig. $2$(b)  and Fig. $2$(c) that  AQCTPCs perform  much better than asymmetric QAG codes.
\begin{figure*}[tbp]
\label{AGCureves}
    \centering
	  \subfloat[ ]{
       \includegraphics[width=0.32\linewidth]{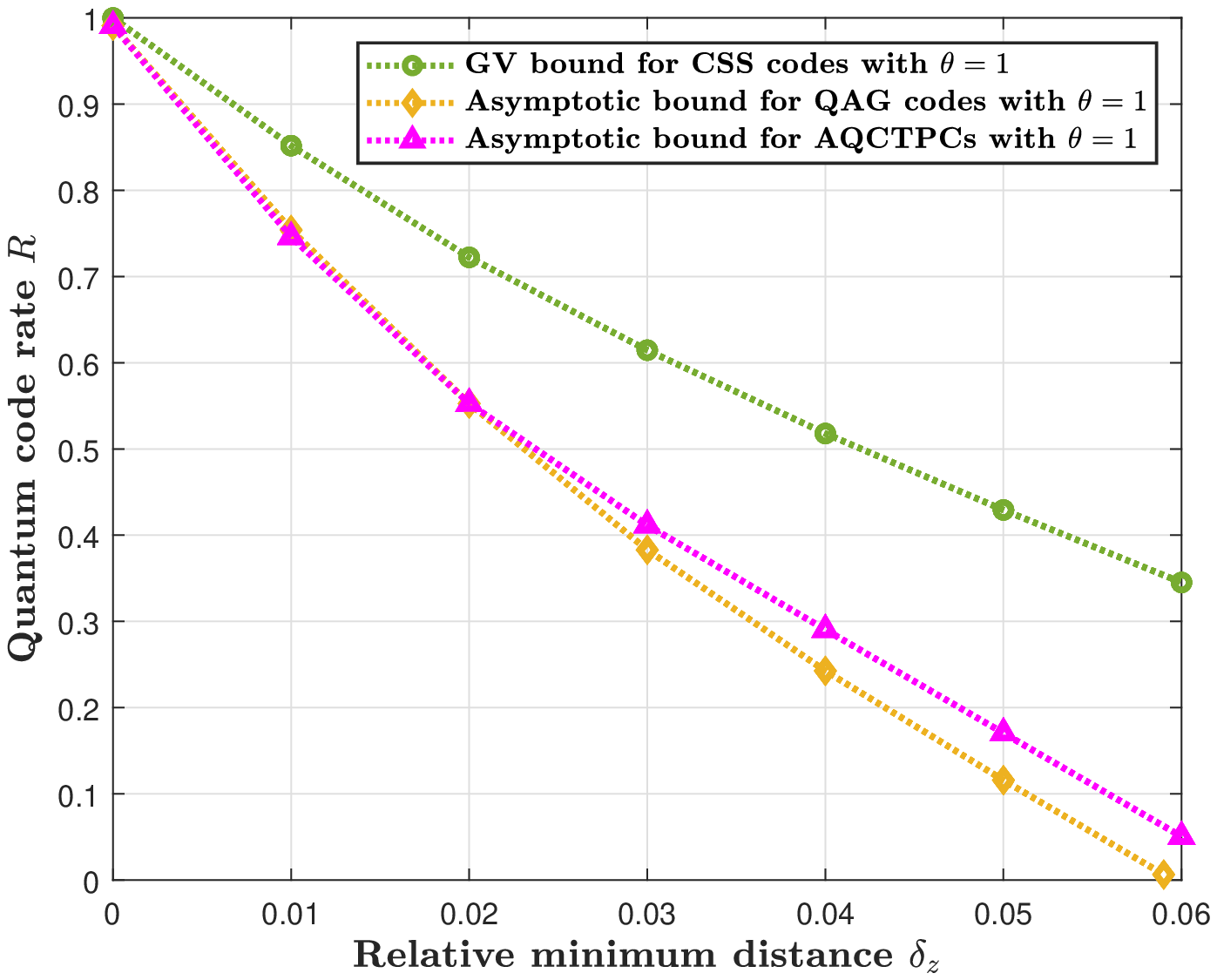}}
    \label{1a}\hfill
	  \subfloat[ ]{
        \includegraphics[width=0.32\linewidth]{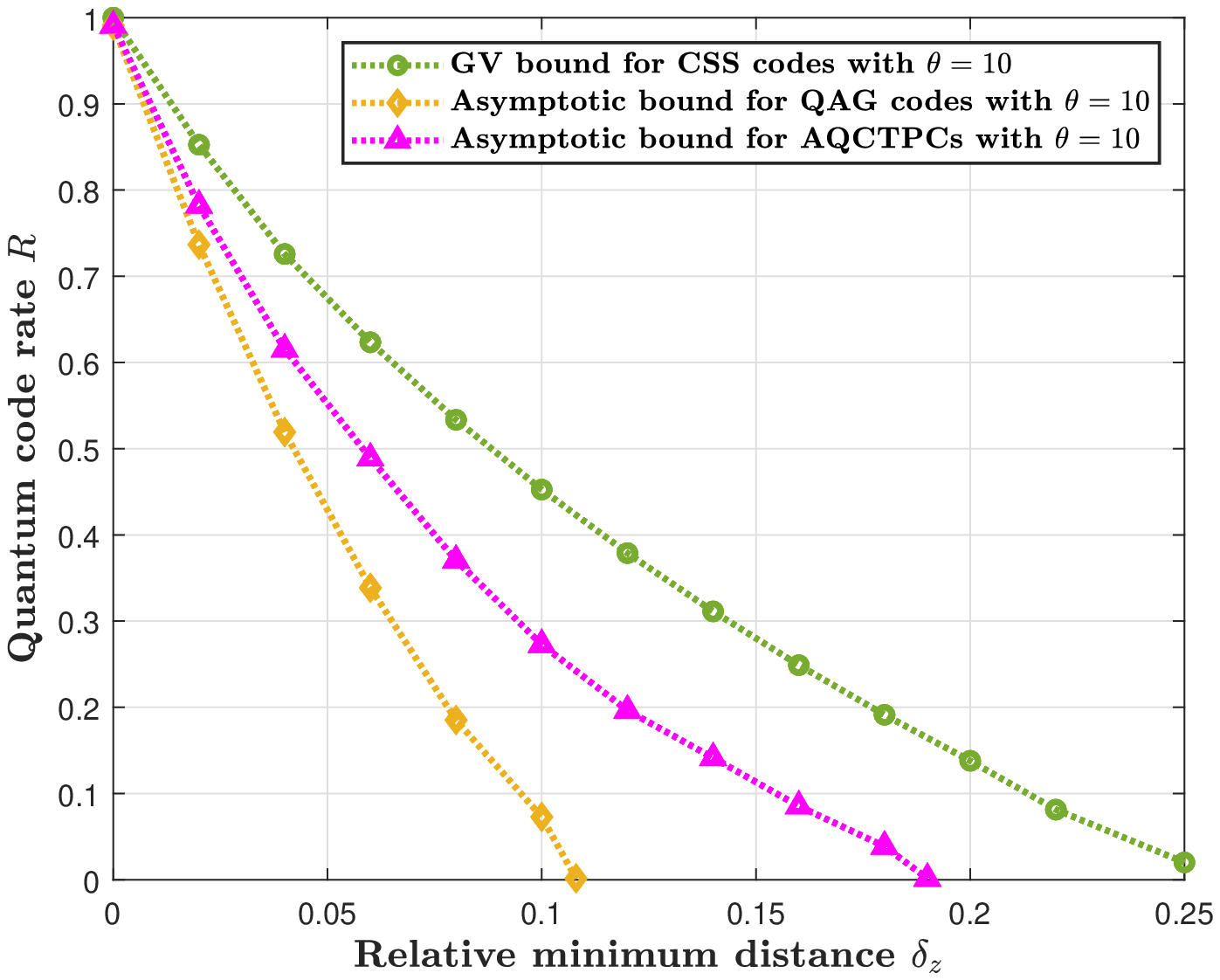}}
    \label{1b} \hfill
	  \subfloat[ ]{
        \includegraphics[width=0.32\linewidth]{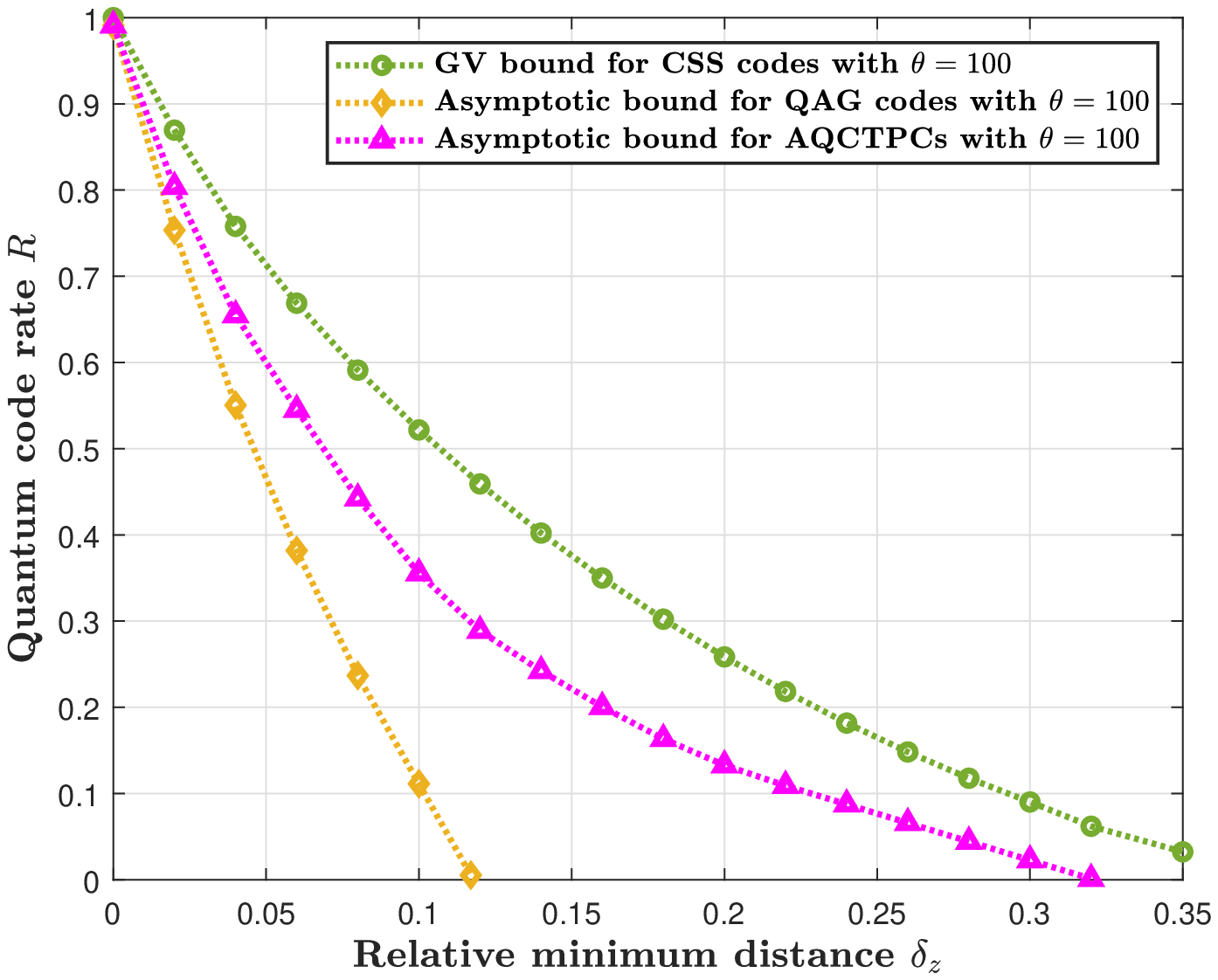}}
    \label{1c}\hfill	
    \caption{The comparison   among the  asymptotic bound for AQCTPCs, the GV bound for CSS codes and the asymptotic bound for asymmetric QAGs. The asymmetry parameter $\theta=d_Z/d_X$ is chosen as $1,10,100$ in (a), (b), and (c), respectively. We use the relative minimum distance $\delta_z=d_Z/N$ as the horizontal axis and use   code rate $R=K/N$ as the  vertical axis. In order to  optimize
 the asymptotic curves for AQCTPCs, we use different constituent code parameters in Ref. \cite{Grassl:codetables} to generate several piecewise asymptotic curves for AQCTPCs and then we joint them together.   }
	  \label{fig1}
\end{figure*}

\section{Conclusions and Discussions}
\label{disandconc}
In this paper, we proposed the construction of asymmetric quantum concatenated and tensor product codes that combine  the classical CCs and TPCs. The CCs correct the $Z$-errors and the TPCs correct the $X$-errors. Compared to concatenation schemes like CQCs and QTPCs, the AQCTPC construction only requires that the outer constituent codes satisfy the dual-containing constraint; the inner constituent codes can be chosen freely. Further, AQCTPCs are highly degenerate codes and, as a result, they passively correct many $X$-errors. To avoid issues with decoding, we present efficient syndrome-based decoding algorithms and show that if the inner and outer constituent codes are efficiently decodable, then the AQCTPC is also efficiently decodable.   Particularly,
 the  inner   decoding complexity of TPCs is significantly reduced
 to $O(n_2)$  in general. Further, we  generalized the AQCTPC concatenation scheme by using GCCs and GTPCs.

To showcase the power of the method, we constructed many state-of-the-art AQCs. Through these constructions, we demonstrate how AQCTPCs can be   superior to QBCH codes or  asymmetric QAG codes as the block length goes to infinity; how they can have better parameters than the binary extension of asymmetric QRS codes; and how varieties of AQCTPCs with a large  $Z$-distance   $d_Z$ can be designed by using some best known linear codes in \cite{Grassl:codetables}. In particular, we constructed a family of AQCTPCs with a  $Z$-distance   $d_Z$ of approximately half the block length,  and meanwhile with dimension   and  $X$-distance   $d_X$ that continue to increase as the block length goes to infinity. If $d_X=2$, we obtain the first family of binary AQCs with the $Z$-distance larger than half the block length.

 Our codes are   practical to quantum communication channels with a large asymmetry and may be used in   fault-tolerant quantum computation to deal with highly biased noise. In the next work,  we may consider the construction and decoding of AQCTPCs by using some other  constituent codes, e.g., the Polar codes.

\section*{Acknowledgment}
The authors would like  to thank the Editor and the referees for their valuable comments that are  helpful to improve the presentation of their article. J. Fan thanks  Prof. Yonghui Li and Prof. Martin Bossert for some earlier communications about tensor product codes.
\bibliographystyle{IEEEtran}
\bibliography{IEEEabrv,tcom}
\ifCLASSOPTIONcaptionsoff
  \newpage
\fi

\end{document}